\documentclass[twocolumn]{aastex6}

\begin{document}
\title{Chlorine Abundances in Cool Stars}

\author{Z. G. Maas}
\affil{Indiana University Bloomington, Astronomy Department, Swain West 319, 727 East Third Street, Bloomington, IN 47405-7105, 
USA}
\email{zmaas@indiana.edu}

\and

\author{C. A. Pilachowski} 
\affil{Indiana University Bloomington, Astronomy Department, Swain West 319, 727 East Third Street, Bloomington, IN 47405-7105, 
USA}
\email{cpilacho@indiana.edu}
\and

\author{K. Hinkle}
\affil{National Optical Astronomy Observatory, P.O. Box 26732, Tucson, AZ 85726, USA}
\email{hinkle@noao.edu}

\begin{abstract}

Chlorine abundances are reported in 15 evolved giants and one M dwarf in the solar neighborhood. The Cl abundance was measured using the vibration-rotation 1-0 P8 line of H$^{35}$Cl at  3.69851 $\mu$m. The high resolution L-band spectra were observed using the Phoenix infrared spectrometer on the Kitt Peak Mayall 4m telescope. The average [$^{35}$Cl/Fe] abundance  in stars with --0.72$<$[Fe/H]$<$0.20 is [$^{35}$Cl/Fe]=(--0.10$\pm$0.15) dex. The mean difference between the [$^{35}$Cl/Fe] ratios measured in our stars and chemical evolution model values is (0.16$\pm$0.15) dex. The [$^{35}$Cl/Ca] ratio has an offset of $\sim$0.35 dex above model predictions suggesting chemical evolution models are under producing Cl at the high metallicity range. Abundances of C, N, O, Si, and Ca were also measured in our spectral region and are consistent with F and G dwarfs. The Cl versus O abundances from our sample match Cl abundances measured in planetary nebula and \ion{H}{2} regions. In one star where both H$^{35}$Cl and H$^{37}$Cl could be measured, a $^{35}$Cl/$^{37}$Cl isotope ratio of 2.2$\pm$0.4 was found, consistent with values found in the Galactic ISM and predicted chemical evolution models.

\end{abstract}


\section{Introduction}
A full knowledge of stellar abundance patterns in a variety of stellar sources is useful to understand the production of each element and the chemical enrichment history of stellar populations. Galactic chemical evolution models rely on theoretical yields which are tested through observations of stellar abundances. Multiple elements have been well studied in galactic evolution, particularly the alpha elements and iron peak elements in Galactic populations. Some odd atomic number light elements are difficult to measure due to their low abundances and spectral features that lie outside optical wavelengths. Thus the light, odd elemental abundance patterns are poorly constrained \citep{nomoto2}. 

This work focuses on chlorine, an odd Z element with two stable isotopes, $^{35}$Cl and $^{37}$Cl. Both chlorine isotopes are formed during hydrostatic and explosive oxygen burning phases \citep{woosley}, although the explosive oxygen burning phase during supernova produces a higher yield than hydrostatic burning \citep{woosley}. A model for a 25 M$_{\odot}$ supernova explosion shows the $^{35}$Cl abundance generated from hydrostatic burning is 4.82 x 10$^{-4}$ M$_{\odot}$ while explosive oxygen burning generates an abundance of 7 x 10$^{-4}$ M$_{\odot}$ \citep{woosley}. $^{35}$Cl is thought to be primarily produced when $^{34}$S captures a proton. The other isotope, $^{37}$Cl is thought to be primarily produced from radioactive decay of $^{37}$Ar, and can be created during neon burning \citep{woosley}.  Models of chlorine production in both core collapse supernova (CCSNe) \citep{woosley, kobayashi6, nomoto, kobayashi11} and Type Ia supernova \citep{travaglio} show chlorine yields vary as a function of mass and metallicity of the progenitor star. These yields have led Galactic enrichment models to predict constant [Cl/Fe]\footnote{[A/B] $\equiv$ log(N$_{A}$/N$_{B}$)$_{star}$ - log(N$_{A}$/N$_{B}$)$_{\odot}$ } ratios over a range of metallicities \citep{kobayashi11}. 

Measurement of the chlorine abundances in stellar atmospheres is extremely difficult and little empirical data are available to compare with Cl evolution models. The solar Cl abundance in particular highlights the difficulties in Cl abundance measurements; no chlorine abundance measurement is possible from the quiet photosphere. An early attempt using near infrared measurements of weak atomic Cl lines provided inconclusive abundance measurements and gave an upper limit on the chlorine abundance\footnote{A(X)=12+log(X/H)}, A(Cl) $\leq$ 5.5 \citep{lambert}. 

Chlorine can also be measured in molecular form. Hydrogen chloride (HCl) molecular vibrational-rotational lines are found in the L-band spectrum of stars. However, the low dissociation potential of HCl limits the molecule to lower temperature stellar atmospheres. HCl features in the spectral range 3.633 $\mu$m and 4.166 $\mu$m have been measured in solar sunspot umbrae spectra and resulted in a solar Cl abundance of 5.5$\pm$0.3 \citep{hall}. Chlorine features at x-ray wavelengths can also be present during solar flares; an abundance of A(Cl)=5.75$\pm$0.26 has been measured using the \ion{Cl}{14} line in solar flare spectra \citep{sylwester}. Due to the difficulties in measuring the Cl abundance in stellar atmospheres, \citet{asplund} suggested that a Cl abundance of A(Cl)=5.32$\pm0.07$ derived from nearby \ion{H}{2} regions from \citet{garcia} may be a suitable proxy for the solar abundance. Finally, the  the meteoric value for Cl of 5.25$\pm$0.06 \citep{lodders} may also be a proxy for the Cl abundance in the Sun. 

While no direct chlorine measurements have been reported in the photospheres of stars, forbidden Cl lines, such as features at $\sim$5500$\AA$ in the optical regime, provide chlorine measurements in planetary nebulae and \ion{H}{2} regions. More recent Cl abundance measurements \citep{esteban} in \ion{H}{2} regions were derived by measuring multiple ions of Cl without using ion correction factors. These improved measurements have lowered the Cl abundance measured in nearby \ion{H}{2} regions and allowed the measurement of the Galactic radial Cl abundance gradient. The radial slope finds a Cl abundance at the solar galactocentric radius of A(Cl)=5.05 \citep{esteban}. Both \ion{H}{2} and planetary nebulae surveys have shown that the gradients of Cl and O with galactocentric distance are nearly identical, \citep{esteban} for \ion{H}{2} observations and \citep{henry} for  PN observations, implying Cl and O production are highly correlated. Other planetary nebula studies used Cl as a proxy for the iron abundance, consistent with current models of Cl production \citep{delgado}. 

Millimeter and submillimeter HCl emission features have been measured in the interstellar medium and provide tests of the $^{35}$Cl/$^{37}$Cl ratio in both Galactic and extragalactic sources. The solar system  $^{35}$Cl/$^{37}$Cl ratio of 3.13 was found from meteoric Cl isotope abundances \citep{lodders}. Measurements of proto-planetary cores have found a $^{35}$Cl/$^{37}$Cl ratio of 3.2 $\pm$0.1 \citep{kama} matching solar values. A survey of 27 star forming regions, molecular clouds, and the circumstellar envelops of evolved stars, observed with the Caltech Submillimeter Observatory, found $^{35}$Cl/$^{37} $Cl ratios that varied between 1 and 5 (uncertainties ranged from 0.3 to 1.0) with the majority of sources having a ratio between 1.1 and 2.5 \citep{peng}. Chlorine has also been detected in extragalactic sources; recently, H$_{2}$Cl$^{+}$ has been detected in a lensed blazar with a measured $^{35}$Cl/$^{37}$Cl ratio of 3.1$^{0.3}_{0.2}$ at a redshift of 0.89 \citep{muller}.  

We report here our Cl abundances measured in stars using HCl molecular lines at infrared wavelengths. Previous infrared spectroscopic studies found HCl molecular lines in the 4 $\mu$m wavelength regime in the atmospheres of cool AGB stars \citep{lebzelter}. This paper describes the observations, source selection, and data reduction in section \ref{sec::reduction}. Section \ref{sec::abund} describes how the line-list was constructed, the spectral synthesis implementation, and the uncertainties in our abundance measurements. Section \ref{sec::discussion} discusses the atmospheric parameters necessary for HCl to form in a solar atmosphere, the Cl isotopic abundance, and comparisons to theoretical Cl enrichment models.

\section{Observations and Data Reduction}\label{sec::reduction}

\begin{figure*}[t]
\centering
\includegraphics[trim=4cm 0cm 4cm 0cm, clip=True, scale=.34]{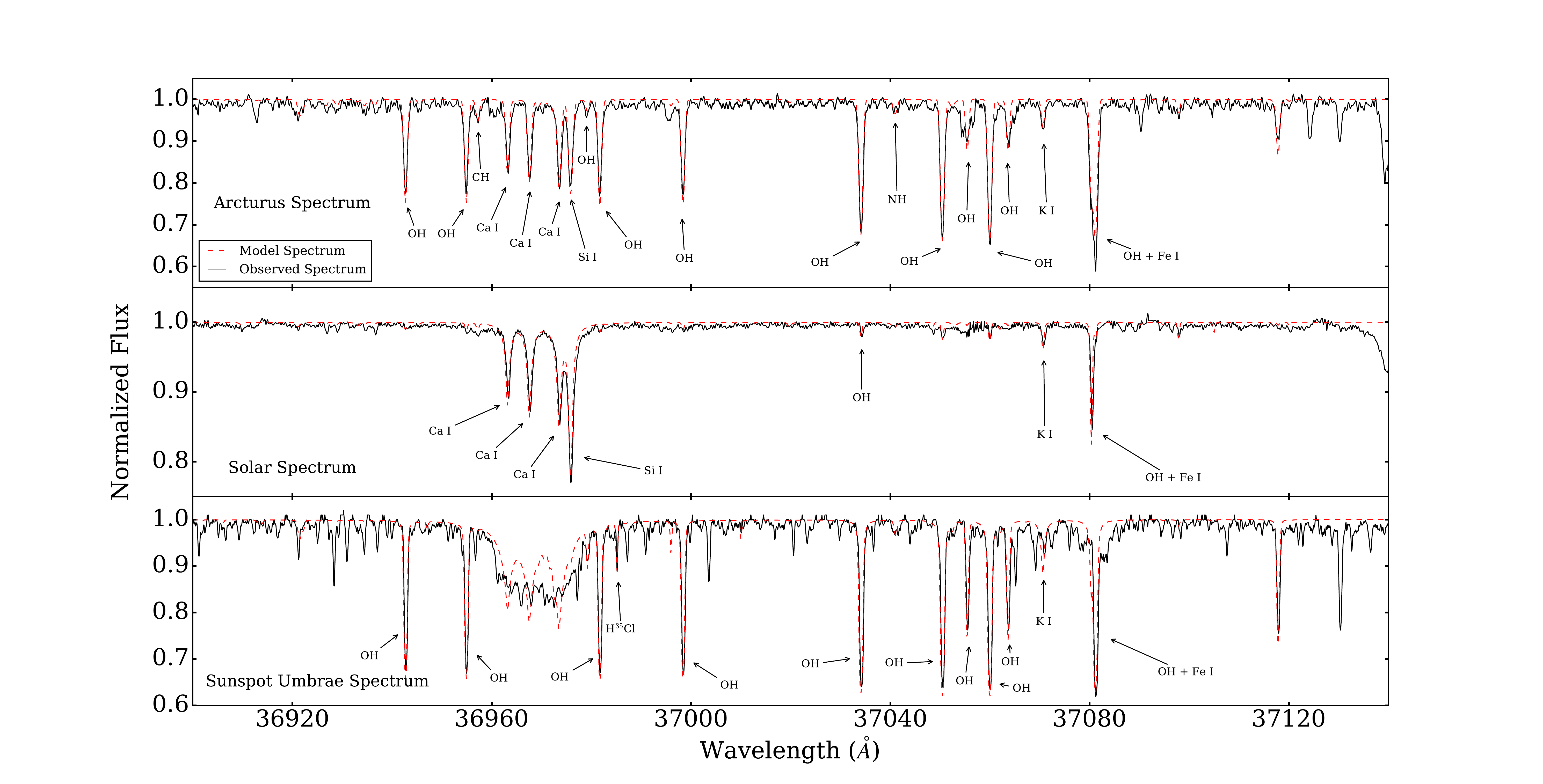} 
\caption{Synthetic spectrum fits (dashed lines) to the Arcturus spectrum from \citet{hinkle} (top panel), solar spectrum from \citet{livingston} (middle panel), and sunspot spectrum from \citet{wallace2} (bottom panel). Zeeman splitting due to the sunspot's magnetic field affects the lines of Si, K, and Ca in the bottom panel. \label{fig::loggffits}}
\end{figure*}

The HCl vibration-rotation fundamental is located in the 3 $\mu$m region where the telluric spectrum can be strong.  Due to the large rotational constant (B$_{e}$ $\approx$10.6 cm$^{-1}$) for the $^1\Sigma$ HCl vibration-rotation transition, the spectral lines are widely spaced.  To select optimum regions to observe we examined both telluric and Arcturus spectra \citep{hinkle} at the wavelength of each H$^{35}$Cl line.  HCl does not appear in the spectrum of Arcturus so this process provided a list of the regions with the fewest blends.  The 1-0 P8 H$^{35}$Cl transition at 3.69851 $\mu$m was selected for observation. The 3.7 $\mu$m spectral region is show in Figure \ref{fig::loggffits} for the Arcturus, solar, and sunspot umbrae spectra. These spectra were all retrieved from NOAO atlases; the solar spectrum was taken from \citet{livingston}, the Arcturus spectrum from \citet{hinkle}, and the the sunspot umbrae spectra from \citet{wallace2}. The H$^{35}$Cl molecular feature is only observable in low temperature stars.  In the Figure \ref{fig::loggffits} H$^{35}$Cl is present in the cool sunspot spectrum but not in the Arcturus atlas or solar atlas.

The program stars were observed with the Phoenix infrared spectrometer \citep{hinkle_et_al_1998} at the f/16 focus of the KPNO Mayall 4 meter telescope in  2014 December and 2015 June. The 0.7 arcsecond four pixel slit was used resulting in a spectral resolution of $\sim$50,000. On 2014 December 7 and 2014 December 8 and 2015 June 7 and 8 the 3.6940 $\mu$m to 3.7110 $\mu$m region of echelle order 15 was selected with a narrow band order sorting filter.
On 2014 December 9 and 10 the spectral range was  slightly shifted, 3.6900 $\mu$m to 3.7070 $\mu$m. 

\begin{deluxetable*}{l l l l c c}
\tablewidth{0pt} 
\tablecaption{Summary of Phoenix Observations \label{table::obslog}} 
\tablehead{\colhead{Star} & \colhead{HD} & \colhead{UT Time} & \colhead{Spectral} & \colhead{K$_{s}$} & \colhead{S/N} \\
\colhead{} & \colhead{Number} & \colhead{And Date} & \colhead{Type} & \colhead{(Mag)} }
\startdata
BD+68 946 & - & 7:31 2015 June 7 & M3.5 V &   4.60 & 70 \\
$\beta$ And& 6860 & 5:05 2014 December 10 & M6 III &   --1.85 & 100 \\
RZ Ari & 18191 & 3:31 2014 December 9 & M0 III &   --0.87 & 180 \\
$\alpha$ Cet & 18884 & 5:08 2014 December 8 & M1.5 III &   --1.82 & 130 \\
Omicron Ori & 30959 & 6:38 2014 December 10 & M3.2 III &   --0.66 & 120 \\
V1261 Ori  & 35155 & 7:50 2014 December 10 & S4 &   2.14 & 130 \\
$\mu$ Gem  & 44478 & 8:35 2014 December 10 & M3 III &   --1.86 & 130 \\
V613 Mon & 49368 & 9:27 2014 December 9 & S5 &   2.46 & 140 \\
27 Cnc & 71250 & 10:39 2014 December 10 & M3 III &   0.58 & 130 \\
$\delta$ Vir & 112300 & 3:35 2015 June 8 & M3 III & --1.19 & 160 \\
83 UMa & 119228 & 4:16 2015 June 7 & M2 III, Ba 0.5 & 0.34 & 160 \\
AW CVn & 120933 & 4:37 2015 June 7 & K5 III & --0.01 & 170 \\
52 Boo & 138481 & 6:06 2015 June 7 & K5 III, Ba 0.5 & 1.21 & 200 \\
\nodata & 147923 & 6:38 2015 June 7 & M2 [S?] & 3.46 & 130 \\
$\lambda$ Aqr & 216386 & 1:36 2014 December 10 & M2 III & --0.67 & 130 \\
30 Psc (YY Psc) & 224935 & 1:44 2014 December 10 & M3 III & --0.40 & 130 \\
 \tableline
$\nu$ Psc & 10380 & 4:09 2014 December 7 & K3 III & 1.36 &  110 \\
\nodata & 10824 & 4:27 2014 December 7 & K4 III & 1.74 & 120 \\
14 Tri & 15656 & 4:46 2014 December 7 & K5 III & 1.63 & 80 \\
\nodata & 20468 & 5:36 2014 December 7 & K2 II & 1.46 & 110 \\
\nodata & 20644 & 5:46 2014 December 7 & K4 III & 0.88 & 130 \\
23 Eri & 23413 & 6:05 2014 December 7 & K4 III & 2.40 & 100 \\
\nodata & 29065 & 6:22 2014 December 7 & K4 III & 1.82 & 100 \\
\nodata & 52960 & 9:31 2014 December 7 & K3 III & 2.04 & 140 \\
$\gamma$ CMi & 58972 & 9:44 2014 December 7 & K3 III & 0.99 & 100 \\
$\upsilon$ Gem & 60522 & 9:53 2014 December 7 & M0 III & 0.23 & 130 \\
g Gem & 62721 & 10:03 2014 December 7 & K4 III & 1.24 & 110 \\
\nodata & 88230 & 12:15 2014 December 7 & K8 V & 2.96 & 130 \\
\nodata & 218792 & 2:20 2014 December 7 & K4 III & 2.66 & 100 \\
22 Psc & 223719 & 2:06 2014 December 7 & K4 III & 2.04 & 130 \\
3 Cet & 225212 & 2:32 2014 December 7 & K3 Iab & 1.40 & 130 \\
\enddata
\end{deluxetable*}

There is considerable thermal background emission at 3.7 $\mu$m.   To remain within the linear response of the detector single integrations were limited to $\sim$150s. Depending on the magnitude of the individual target, exposure times were between 1 to 150 seconds.  The noise contributed by thermal emission from background radiation limited the stellar sample to bright infrared sources. All the stars observed had K$<$5.  Spectral type K and M, Population I, stars were targeted. A full list of stars that showed chlorine feature are displayed in Table \ref{table::obslog}. The second portion of this table shows a list of stars that were analyzed but did not present HCl features in their spectra. They are included to extend our analysis over large temperature ranges. Stars with effective temperatures between $\sim$3300 K and $\sim$ 4300 K were chosen as candidate stars for the observations.

The spectral region covered does not contain enough spectral features to determine atmospheric parameters for the sample stars. While numerous stellar OH lines are present in this wavelength range these lines do not cover a large enough range of excitation potential suitable to set atmospheric parameters for the sample stars. Thus we selected stars with atmospheric parameters available within the literature.

Standard observing procedures were used with the star nodded between two slit positions. Each program star was observed close to zenith with an airmass less than 1.4, most between 1.0 and 1.1. Additional standard stars to be used for telluric line removal were observed with air masses similar to target stars. Flat and dark calibration images were taken at the beginning of each night. 

Data reduction was accomplished using the IRAF software suite\footnote{IRAF is distributed by the National Optical Astronomy Observatory, which is operated by the Association of Universities for Research in Astronomy, Inc., under cooperative agreement with the National Science Foundation.}. The standard near-IR reduction process discussed by \citet{joyce_1992} was followed. After each image was trimmed, the dark images were combined using a median filter with a sigclip rejection algorithm. The flats were median combined with an avsigclip rejection algorithm. The combined dark image was subtracted from the combined flat image. Each set of stellar observations had been taken in groups of four with two different slit positions using the sequence "abba". Using the IRAF task imarith, adjacent observations were subtracted from one another to remove sky, dark, and common detector blemishes. Each image was then divided by the dark subtracted, normalized flat field image.

\begin{figure*}
\epsscale{1} 
\centering
\includegraphics[trim=5cm 0cm 5cm 0cm, clip=True, scale=.32]{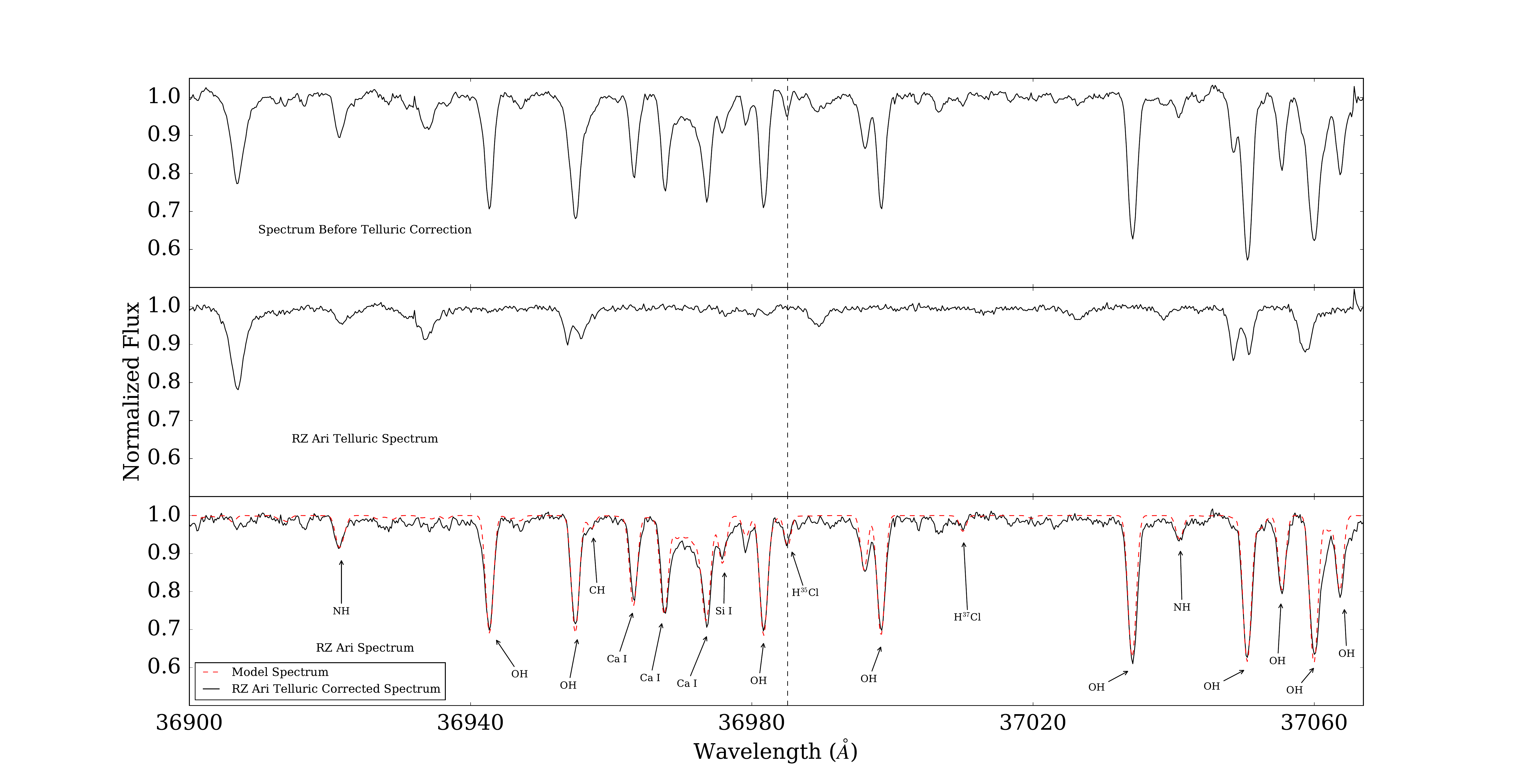} 
\caption{Top Panel: The spectrum before telluric correction is shown for the star RZ Ari. Midde Panel: This panel shows the telluric lines present in the spectra. Bottom Panel: The final spectra (solid line) and best fit (dashed line) for RZ Ari is shown. Note the detection of H$^{37}$Cl feature at 37010 $\AA$. \label{fig::exspec2}} 
\end{figure*}

The resulting images were extracted using the IRAF task apall and multiple observations were combined using the IRAF task scombine. The images from the December 2014 observing runs were either median or sum combined depending on the cosmic ray and bad pixel contamination. After the combination process, each spectrum was normalize. Telluric lines were removed using the IRAF task 'telluric.' This correction was done using standard stars observed each night. The telluric contamination in the spectral region is minimal, shown in Figure \ref{fig::exspec2}. Each telluric correction was done with a standard star that most closely matched the air mass of the observed star. Finally, the exposures were wavelength calibrated using stellar lines identified in the Arcturus atlas and in each of the stars.

\section{Abundance Analysis}\label{sec::abund}
\subsection{Line List Construction}

The line list, shown in Table \ref{table:: linelist}, was derived from multiple sources sources, the Kurucz atomic line database\footnote{http://kurucz.harvard.edu/atoms.html},  \citep{brooke} for OH lines\footnote{linelist from http://www.as.utexas.edu/$\sim$chris/lab.html},  \citep{brookenh,brookenh2} for NH lines, and the HITRAN database \citep{rothman}. The \citet{brooke} OH log gf values for lines in our spectral range are consistent with those in the Kurucz molecular line data base. The atomic line parameters were adopted from the Kurucz database and CH values were from \citet{masseron_ch} (values listed in the Kurucz database). The oscillator strength (f-value) for the HCl features was calculated from the Einstein A coefficient recorded in the HITRAN database, using Equation \ref{eq::loggf} \citep[][pg.153]{bernath}. 

\begin{equation}
\label{eq::loggf}
f_{J' \leftarrow J''}=\frac{1.4991938}{\nu^{2}/(wavenumber^{2}) }\frac{2J' +1}{2J'' +1} A_{J'\rightarrow J''}/(s)
\end{equation}

The log gf values were tested by fitting high resolution solar photospheric spectrum and  Arcturus spectrum using synthetic spectra. The atmospheric parameters and abundances for each star can be found in Table \ref{table::params}. The solar abundances are adopted from \citet{asplund} and the Arcturus atmospheric parameters and abundances were adopted from \citet{ramirez} with the exception of the nitrogen abundance which was adopted from \citet{smith} . 

Plane parallel and spherical atmosphere models for the Sun and Arcturus respectively were generated from the MARCS grid \citep{gustafsson} and were interpolated using the program available on the MARCS site\footnote{http://marcs.astro.uu.se/software.php} \citep{masseron}. Dissociation potentials for each molecular species were found the literature and are listed in Table \ref{table:: linelist}; sources include \citet{ruscic} for OH, \citet{martin} for HCl, \citet{espinosa} for NH, and \citet{kumar} for CH. Synthetic spectra were generated using the 2014 version of the MOOG spectral analysis software \citep{sneden}. The synthetic spectra were convolved with a Gaussian to match the line profiles in Arcturus and solar spectra. The best fits to the spectra are shown in Figure \ref{fig::loggffits}. The astrophysical log gf values found for the molecular lines are consistent with the theoretical values with the exception of two OH lines at 37050.462 $\AA$ and 37059.961 $\AA$; their log gf values were increased by 0.1 to better fit the Arcturus and solar spectra. The log gf values of the atomic features were on average 0.4 larger than the atomic features in the Kurucz database. Fe I features in the spectra could not be fit simultaneously in both the Arcturus spectra and the solar spectrum, possibly due either to blending weak, unidentified molecular lines or other features, or due to errors in the excitation potential. Our best estimates for the log gf values for the Fe I lines are included in Table \ref{table:: linelist}. 

\begin{deluxetable*}{l c c c c c l}
\tablewidth{0pt} 
\tablecaption{Line List \label{table:: linelist}} 
\tablehead{\colhead{Element} & \colhead{$\lambda_{air}$ } & \colhead{$\nu_{vac}$} & \colhead{$\chi$} & \colhead{log gf} & \colhead{D$_{0}$} & \colhead{Transition} \\
\colhead{} & \colhead{($\AA$)} &\colhead{cm$^{-1}$} & \colhead{(eV)} & \colhead{} & \colhead{(eV)} & \colhead{}}
\startdata
NH   & 36921.287 & 2707.70646 &  1.655 & -3.085 & 3.3693 & 5-4 R3e10 \\
NH   & 36921.494 & 2707.69132 &  1.655 & -3.012 & 3.3693 & 5-4 R1e12 \\
NH   & 36921.506 & 2707.69041 &  1.655 & -3.048 & 3.3693 & 5-4 R2f11 \\
OH & 36942.692 & 2706.10656 & 1.247 & --3.859 &  4.4130 & 3-2	P2f12.5 \\
OH & 36954.896 &  2705.21496 & 1.248 & --3.859 &  4.4130 & 3-2	P2e12.5 \\
CH & 36957.174 & 2705.04229 & 0.3580 & --3.570 &  3.4948 & 2-1	R1e3.5 \\
Ca I &  36963.164 & 2704.605 & 5.5637 & 	1.15 & \nodata & 3s$^4$f$^4$F$^{0}_{2}$ - 3s$^4$g$^5$G$^{0}_{3}$   \\
Ca I & 36967.510	&   2704.286 &5.5637  & 1.25  & 	\nodata & 3s$^4$f$^4$F$^{0}_{3}$ - 3s$^4$g$^5$G$^{0}_{4}$ \\	
Ca I & 36967.866	&  2704.260 &5.5637 & 0.10   &	\nodata & 3s$^4$f$^4$F$^{0}_{3}$ - 3s$^4$g$^5$G$^{0}_{3}$\\	
NH   & 36969.559 & 2704.17095 &  0.264 & -4.723 & 3.3693 & 1-0 P1e12 \\
NH   & 36970.709 & 2704.08687 &  0.264 & -4.763 & 3.3693 & 1-0 P2f11 \\
NH   & 36971.829 & 2704.00489 &  0.264 & -4.803 & 3.3693 & 1-0 P3e10 \\
Ca I & 36973.485	&   2703.849  &5.5637 &	1.34 & \nodata &  3s$^4$f$^4$F$^{0}_{4}$ - 3s$^4$g$^5$G$^{0}_{5}$\\
Ca I & 36973.909	& 2703.818 & 5.5637 &	0.10 &	\nodata & 3s$^4$f$^4$F$^{0}_{4}$ - 3s$^4$g$^5$G$^{0}_{4}$\\
Si I & 36975.824	&  2703.678   &  7.0692 & 	0.77 & \nodata & 3p$^{3}$s$^{5}$P$_{1}^{0}$ -  3p$^{3}$p$^{5}$D$_{2}^{0}$\\	
OH & 36979.19 &  2703.43514 & 1.727 & --4.514 & 4.413 & 5-4	P2f6.5\\
OH & 36981.688 & 2703.25474&	 1.243 & --3.825	& 4.413 & 3-2	P1e13.5\\
H$^{35}$Cl & 36985.103 &2703.0074 & 0.0929 & 	--4.178 & 4.432 & 1-0 P8 \\
OH & 36998.379 &	 2702.03348 & 1.244	& --3.825& 4.413	 & 3-2	P2f13.5\\
H$^{37}$Cl & 37010.023 &2701.1874& 0.0929 &  --4.178 & 4.432 & 1-0 P8 \\
OH & 37034.102 & 2699.42489 & 1.030	& --3.782 &	4.413 & 2-1	P2f15.5\\
NH   & 37040.866 & 2698.96521 &  1.619 & -3.148 & 3.3693 & 5-4 R3e9 \\
NH   & 37041.139 & 2698.94526 &  1.619 & -3.108 & 3.3693 & 5-4 R2f10 \\
NH   & 37041.176 & 2698.94261 &  1.618 & -3.069 & 3.3693 & 5-4 R1e11\\
OH & 37050.462 & 2698.23581 & 1.031	 & ---3.682 &  4.413	 & 2-1	P1e15.5 \\
OH & 37055.419 &	 2697.87443 & 1.492  & --4.069 &	4.413 & 4-3	P2f9.5\\
OH & 37059.961 &	 2697.544 &1.026  & --3.655	 &	4.413 &	2-1	P1e16.5\\
OH & 37063.156 & 2697.27596 & 1.492	 & -4.069	 &	4.413	& 4-3	P2e9.5\\
K I & 37070.611 & 2696.765  & 3.3969	 & 0.82	 & \nodata & 2p$^{5}$P$_{0.5}^{0}$ - 2d$^{4}$D$_{1.5}^{0}$  \\
Fe I & 37080.346	& 2696.057 & 5.0674	 &--0.94	 & \nodata & 3p$^4$yD$_{3}^{0}$ - 3d$^{3}$P$^{0}_{3}$ \\
OH & 37081.155 & 2696.00032& 1.028	 & --3.755	 & 4.413 & 2-1	P2f16.5 \\
OH   & 37117.883 & 2693.33812&	1.487 & --4.024 &	 4.413 & 4-3	P1e10.5\\
\enddata
\end{deluxetable*}

The H$^{35}$Cl molecular line is not present in either Arcturus or the solar photosphere due to the high photospheric temperatures, but is found in sunspot umbrae spectra. To fit the sunspot spectrum from \citet{wallace2}, we used a similar methodology as \citet{maiorca} to determine the effective temperature of the sunspot. We generated multiple atmospheric models with different temperatures and found an effective temperature of 3900 K allowed the best fit to the OH lines. We found an abundance of A($^{35}$Cl)=5.31 $\pm$0.12  best fit the spectral feature.  An uncertainty of 0.10 dex was found from the model temperature by fitting the Cl feature with models at 4000 K and 3800 K and the uncertainty on the fit was determined to be 0.06 dex. However, because we do not attempt to model the complex radiative transfer in sunspots, we chose to adopt a solar chlorine abundance of A(Cl)=5.25 based on the meteoric Cl abundance \citep{lodders}.

\subsection{Model Atmosphere Parameters}

\begin{figure*}[t!]
\epsscale{1} 
\centering
\includegraphics[trim=0cm 0cm 0cm 0cm, clip=True, scale=.29]{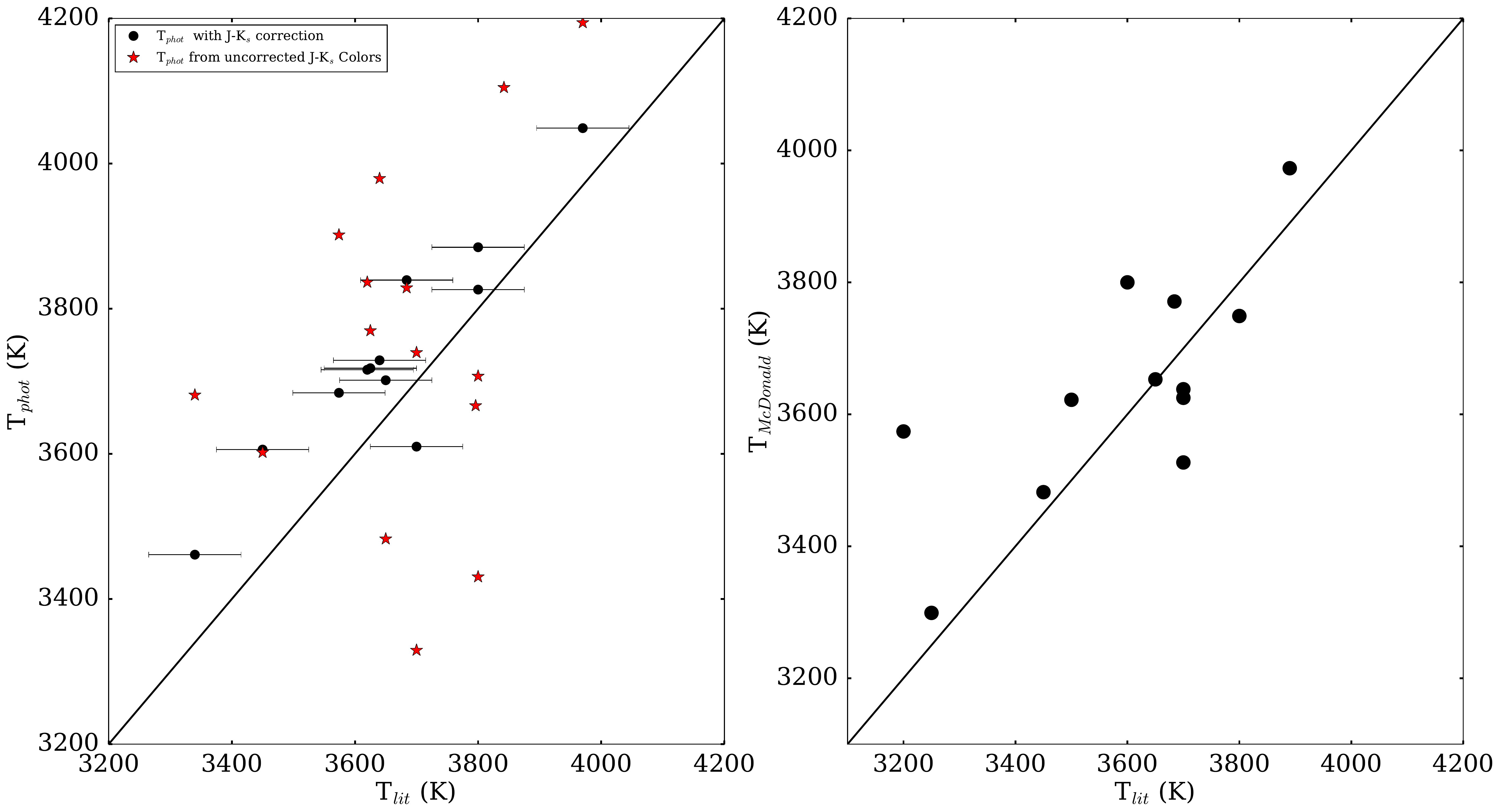} 
\caption{Left Panel: Temperatures derived from the literature for the 15 giant stars with chlorine abundance measurements are plotted against temperatures derived from J-K$_{s}$ colors. The black points represent stars that have been corrected using observed to modeled flux ratios from \citet{mcdonald} while the red stars represent temperatures from raw 2MASS J-K$_{s}$ colors. Error bars corresponding to the $\pm$75 K assigned to the literature temperatures are plotted. A one to one linear line is also plotted for reference. Right Panel: Temperatures from \citet{mcdonald} are compared to temperatures from various sources listed in Table 3. A one to one linear line is plotted for reference.  \label{fig::temp_colors}} 
\end{figure*}

The spectral range is not sufficiently large to derive atmospheric parameters for the target stars. The T$_{eff}$, log g, [Fe/H], and $\xi$ values for each star were adopted from literature sources and can be found in Table \ref{table::params} along with references for each value. Effective temperatures for these stars were mostly determined by photometry (eg. \citealt{smith2}) and spectral energy distribution fitting (e.g. \citealt{mcdonald}). For stars with multiple temperatures cited in the literature, parameters were chosen based on the resulting oxygen abundance from the fit. The majority of the stars in the sample are metal rich and have oxygen abundances near [O/Fe]=0. 

The adopted temperatures for each star were compared with temperatures derived using J-K$_{s}$ colors from 2MASS photometry \citep{skrutskie} and with a color-effective temperature relationship from \citet{gonzalez}. In our sample, 13 of the 15 giants are beyond the J-K$_{s}$  $\geq$ 0.9 limit for the color-temperature relation \citep{gonzalez} and the relation was extrapolated up to colors of 1.22 with most between 0.9 and 1.1.  The V magnitude was not used to minimize the effect of extinction from circumstellar dust. Initial temperatures were compared to literature sources and are shown in Figure \ref{fig::temp_colors} (left panel). The average difference between photometric temperatures and those from the literature is 51 K $\pm$ 239 K. Extinction due to circumstellar dust in the J and K$_{s}$ bands were corrected using observed to model flux ratios from \citet{mcdonald}. Three stars were not included in \citet{mcdonald}, $\beta$ And, $\alpha$ Cet, and $\delta$ Vir. The average correction to the J-band magnitude was 0.01 mags $\pm$  0.10 mags and K$_{s}$-band corrections had an average of 0.005 mags $\pm$ 0.04 mags. The J-band standard deviation is dominated by the star HD 147923 with a correction of 0.28 mags. Without this star, the average correction is $\Delta$J= --0.01 mags $\pm$ 0.06 mags. The temperatures derived from the corrected J-K$_{s}$ colors are shown in the left panel of Figure \ref{fig::temp_colors}. The average difference between the temperatures adopted from the literature and the temperatures calculated from the J-K$_{s}$ color is 76 K $\pm$ 95 K.  

An additional test was performed by comparing temperatures derived by \citet{mcdonald} to those in the literature (see Table 3) for the 12 giants in our sample with measurable Cl abundances, shown in the right panel of Figure \ref{fig::temp_colors}. Some stars in Table 3 have two citations; one that represents a temperature from \citet{mcdonald} and another for the other atmospheric parameters. The temperature from the source associated with the atmospheric parameters is compared to the quoted \citet{mcdonald} temperature in those instances. The average difference between the \citet{mcdonald} temperature and those from other sources is -49 K $\pm$ 143 K. Excluding the outlier of  27 Cnc, which has a difference of -374 K, brings the average down to -19 K $\pm$ 105 K. The corrected J-K$_{s}$ temperatures, temperatures from \citep{mcdonald}, and temperatures from other sources are consistent.

\begin{deluxetable*}{l c c c c c c c c c c}
\tabletypesize{\footnotesize}
\centering
\tablewidth{0pt} 
\tablecaption{Atmospheric Parameters and Abundances \label{table::params}} 
\tablehead{ \colhead{Star} & \colhead{T$_{eff}$ } & \colhead{log g} & \colhead{[Fe/H]} & \colhead{$\xi$ } & \colhead{A(C)} & \colhead{A(N)} & \colhead{A(O)} & \colhead{A(Si)} & \colhead{A($^{35}$Cl)}  & \colhead{A(Ca)} \\
\colhead{} & \colhead{(K)} & \colhead{} & \colhead{} & \colhead{(km s$^-1$)} } 
\startdata
 Solar  & 5780 & 4.40 & 0.00\tablenotemark{a} & 0.75 & 8.43 &  7.83 & 8.69 & 7.51 & 5.13\tablenotemark{q}  & 6.34 \\
 $\alpha$ Boo & 4290 & 1.60 & --0.52 & 1.67  & 8.34 & 7.69 & 8.60 & 7.32 & \nodata  & 5.90 \\
 Sunspot Umbra & 4000 & 4.40 & 0.00 & 0.75 & 8.43 & 7.83 & 8.69 & 7.51 & 5.13\tablenotemark{q}  & 6.34 \\
 \tableline 
 BD+680946\tablenotemark{n} & 3600 & 5.00 & 0.00 & 1.00 & \nodata & \nodata & 8.74 & \nodata & 5.27  & \nodata \\
 HD 6860\tablenotemark{m} & 3842 & 0.90 & --0.20 & 2.20  & 8.33 & 7.96 & 8.62 & 7.09 & 4.90 & 6.11 \\
 HD 18191\tablenotemark{j} & 3340\tablenotemark{g} & 0.30 & --0.25 & 2.45 & 8.33 & 7.99 & 8.47 & 7.36 & 4.82 & 6.43 \\
 HD 18884\tablenotemark{d} & 3796 & 0.91 & --0.45 & 2.00 & 8.33 & 7.63 & 8.60 & 6.97 & 4.86 &  6.25 \\
 HD 30959\tablenotemark{b} & 3450 & 0.80 & --0.15 & 2.95 & 8.38 & 8.09 & 8.56 & 7.41 & 4.88 &  6.13 \\
 HD 35155\tablenotemark{l} & 3650 & 0.80 & --0.72 & 2.30 & 8.30 & 7.64 & 8.45 & 7.04 & 4.57 &  5.91 \\
 HD 44478\tablenotemark{i} & 3640\tablenotemark{g} & 1.00 &  0.00 & 2.45 & 8.43 & 8.03 & 8.83 & 7.43 & 5.07 &  6.37 \\
 HD 49368\tablenotemark{l}  & 3700 & 1.00 & --0.45 & 2.50  & 8.42 & 7.99 & 8.56 & 7.26 & 4.90 &  6.24 \\
 HD 71250\tablenotemark{i} & 3574\tablenotemark{g} & 1.00 &  0.00 & 2.35 & 8.53 & 7.73 & 8.77 & 7.31 & 4.88 &  6.04 \\
 HD 112300\tablenotemark{k} & 3700 & 0.80 &  0.17 & 2.00 & 8.70 & 8.07 & 8.96 & 7.23 & 5.27 &  6.44 \\
 HD 119228\tablenotemark{j} & 3684 & 1.00 &  0.00 & 1.80 & 8.69 & 7.94 & 8.89 & 7.42 & 5.05 &  6.32 \\
 HD 120933\tablenotemark{j} & 3625\tablenotemark{g} & 0.98 & --0.09 & 2.00 & 8.47 & 7.71 & 8.65 & 7.08 & 4.88 &  6.00 \\
 HD 138481\tablenotemark{l} & 3970\tablenotemark{g} & 1.60 &  0.20 & 2.40 & 8.90 & 8.18 & 9.09 & 7.32 & 5.50 & 6.29 \\
 HD 147923\tablenotemark{l} & 3800\tablenotemark{g} & 0.80 & --0.19 & 2.20 & 8.44 & 8.19 & 8.64 & 7.25 & 5.11 & 6.25 \\
 HD 216386\tablenotemark{i} & 3800\tablenotemark{o} & 1.00 &  0.05 & 2.10 & 8.61 & 8.17 & 8.92 & 7.26 & 5.38 &  6.21\\
 HD 224935\tablenotemark{i} & 3620\tablenotemark{g} & 1.20 &  0.05 & 2.75 & 8.89 & 8.17 & 8.99 & 7.47 & 5.07 & 6.34 \\
 \tableline 
 HD 10380\tablenotemark{h} & 4110 & 1.93 & --0.17 & 2.10 & 8.47 & 7.83 & 8.67 & 7.22 & \nodata &  6.08 \\
 HD 10824\tablenotemark{h} & 3940 & 1.66 & --0.18 & 2.20 & 8.47 & 7.92 &8.67 & 7.35 & \nodata &  6.18 \\
 HD 15656\tablenotemark{h} & 3990 & 1.65 & --0.21 & 2.20 & 8.51 & 7.70 & 8.70 & 7.30 & \nodata &  6.16 \\
 HD 20468\tablenotemark{n} & 4325 & 1.42 & --0.17 & 3.10\tablenotemark{e} & 8.43 & 7.78 & 8.60 & 6.97 & \nodata & 6.07 \\
 HD 20644\tablenotemark{h} & 3980 & 1.56 & --0.31 & 2.60 & 8.36 & 7.88 &  8.49 & 7.36 & \nodata &  6.24 \\
 HD 23413\tablenotemark{h} & 4050 & 1.74 & --0.23 & 2.10 & 8.47 & 7.84 & 8.66 & 7.44 & \nodata & 6.23 \\
 HD 29065\tablenotemark{h} & 3990 & 1.77 & --0.35 & 2.30 & 8.08 & 8.02 & 8.44 & 7.44 & \nodata &  6.13 \\
 HD 52960\tablenotemark{c} & 4150 & 1.80 & --0.09 & 2.00 & 8.68 & 8.02	& 8.81 & 7.49 & \nodata &  6.28 \\
 HD 58972\tablenotemark{h} & 4000 & 1.82 & --0.37 & 2.20 & 8.27 &  7.87	& 8.49 & 7.17 & \nodata &  5.97 \\
 HD 60522\tablenotemark{c} & 3940\tablenotemark{g}& 1.90 & --0.37 & 2.60 & 8.51 & 7.85 &	8.67 & 7.43 & \nodata & 6.25 \\
 HD 62721\tablenotemark{h} & 3940 & 1.67 & --0.27 & 1.90 & 8.47 &  7.61	& 8.65 & 7.33 & \nodata &  6.25 \\
 HD 88230\tablenotemark{p} & 3970 & 4.51 & --0.05 & 1.00 & \nodata & 8.59	& 7.85 & \nodata & \nodata & 	\nodata \\
 HD 218792\tablenotemark{h} & 4160 & 2.06 & --0.14 & 2.20 & 8.54 & 7.80 & 8.72 & 7.52 & \nodata &  6.21 \\
 HD 223719\tablenotemark{h} & 3960 & 1.63 & --0.24 & 2.40 & 8.37 & 7.93 &	8.59 & 7.30 & \nodata 	& 6.07 \\
 HD 225212\tablenotemark{f} & 4125 & 0.90 & 0.08 & 2.36 & 8.68 &  8.16 &	8.82 & 7.24	& \nodata 	& 6.42 \\ 
\enddata
\tablecomments{A footnote on the star name (e.g. HD225212$^{f}$) indicates all parameters were take from the source indicated by the footnote. Any additional footnotes over a single parameter (such as effective temperature) indicate only that one parameter was taken by the source represented by the footnote and the rest of the atmospheric parameters came from the source indicated on the star name.}
\tablenotetext{a}{This paper adopts A(Fe)$|_{\odot}$=7.50}
\tablenotetext{b}{\citet{cenarro} } \tablenotetext{c}{\citet{hekker}} \tablenotetext{d}{\citet{jofre}}
\tablenotetext{e}{\citet{luck95}}
\tablenotetext{f}{\citet{luck}}
\tablenotetext{g}{\citet{mcdonald}}
\tablenotetext{h}{\citet{mcwilliam}}
\tablenotetext{i}{\citet{nault}}
\tablenotetext{j}{\citet{prugniel}}
\tablenotetext{k}{\citet{sheffield}}
\tablenotetext{l}{\citet{smith2}}
\tablenotetext{m}{\citet{smith}}
\tablenotetext{n}{\citet{soubiran}}
\tablenotetext{o}{\citet{tsuji}}
\tablenotetext{p}{\citet{woolf}}
\tablenotetext{q}{A($^{35}$Cl)$|_{\odot}$=5.13, A(Cl)$|_{\odot}$=5.25}
\end{deluxetable*}

The AAVSO database\footnote{From https://www.aavso.org/} was examined to determine whether our cool red giant stars were included. Stars showing significant variability in the AAVSO database include RZ Ari, $\lambda$ Aqr, AW CVn, 83 UMa, $\beta$ And, $\mu$ Gem, and V613 Mon. The least variable of this group is AW CVn with an amplitude of 0.028 mags in the V-band \citep{tabur}. Other studies have shown V1261 Ori with a V-band amplitude of 0.24 mags, an average V-band magnitude of 6.77, a period of 624 days, and an observed minimum on JD 2454552  \citep{gromadzki}. Two K-band measurements of K=2.03 mags on JD 2449653.5 \citep{chen} and K=2.138 on JD 2451108.825 \citep{phillips} can be used to determine a V-K index and estimate the star's effective temperature range due to variability. The range of V-K for this star gives an effective temperature range 3527 K to 3594 K, a range less than 70 K \citep{worthey}. 

Other work on M-giant pulsations shows that the amplitude of variability decreases at longer wavelengths, \citet{percy} finds that the slope of the decrease in V can be represented by $\Delta$V$\sim$1.8 $\pm$ 0.2$\Delta$R and the slope of R is $\Delta$R$\sim$1.8 $\pm$ 0.2$\Delta$I for M-giants. Extrapolating this trend would imply that the K-band amplitude decreases by a factor of 7.2 from the V-band to the K-band. For example RZ Ari, another moderately variable star in our sample, has a maximum V-band amplitude of $\sim$0.25 and therefore would have an approximate K-band amplitude of 0.035. This roughly corresponds to a temperature range of 3143 K to 3273 K, with an associated uncertainty of $\pm$65 K \citep{worthey}. Therefore, even in stars without measured K variability, their V-band variability can be used to characterize the uncertainty in their temperature. In the cases of our two most variable stars, RZ Ari and V613 Mon, this temperature uncertainty is only on the order of $\sim$50 K. 

\begin{deluxetable*}{c c c c c c c c}
\tabletypesize{\footnotesize}
\centering
\tablewidth{0pt} 
\tablecaption{Abundance Uncertainties \label{table::uncertainties}} 
\tablehead{  \colhead{ } & \colhead{$\delta$ A(C)} &  \colhead{$\delta$ A(N)}  &\colhead{$\delta$ A(O)} & \colhead{$\delta$ A(Si)} & \colhead{$\delta$ A(Cl)}  & \colhead{$\delta$ A(Ca)}} 
\startdata
 Fit  &  0.13 & 0.08 & 0.07	& 0.11 & 0.05 & 0.06  \\
 $\delta$ T$_{eff}$=$\pm$75 K  & 0.05 &  0.07 & 0.08 & 0.09 & 0.16  & 0.03 \\
 $\delta$(log g) =$\pm$0.5  & 0.07 &	0.04 & 0.05	 & 0.13	& 0.03 &	0.04 \\
 $\delta$[Fe/H] =$\pm$0.2  & 0.08 & 0.12 & 0.12 & 0.06 & 0.15 &  0.04 \\
 $\delta$ $\xi$=$\pm$0.5 (km s$^{-1}$) & 0.02 & 0.02 & 0.06 & 0.06 & 0.03 & 0.07 \\
 Total Uncertainty & 0.18 & 0.17 & 0.18 & 0.21 & 0.23 &  0.15 \\
\enddata
\end{deluxetable*}

Due to the variety of sources used for atmospheric parameters and uncertainty in the $\delta$(V-K) value, we adopt an uncertainty in temperature of $\pm$75 K. The adopted uncertainty in log g is $\pm$0.5, in [Fe/H] is $\pm$0.2 and in microturbulence ($\xi$) is $\pm$0.5 km s$^{-1}$. A conservative log g error was chosen because literature sources derived log g using a variety of methods. Log g could be derived using the effective temperature, assumed mass, and luminosity of a star (e.g. \citealt{smith2}), analysis of atomic lines in the spectrum (e.g. \citealt{sheffield}), or from models of stellar evolution (e.g. \citealt{tsuji}), and not all sources have assigned uncertainties. The uncertainty on metallicity was set to reflect the uncertainty from multiple sources. For example, uncertainties derived from \citep{smith2} are on average --0.13 dex for (Fe/H) ratios using $\alpha$ Tau as the reference star. 

\subsection{Spectral Synthesis Results} \label{subsec::carbonfitting}

\begin{figure*}[b!]
\epsscale{1} 
\centering
\includegraphics[trim=0cm 0cm 0cm 0cm, scale=.35, clip=True]{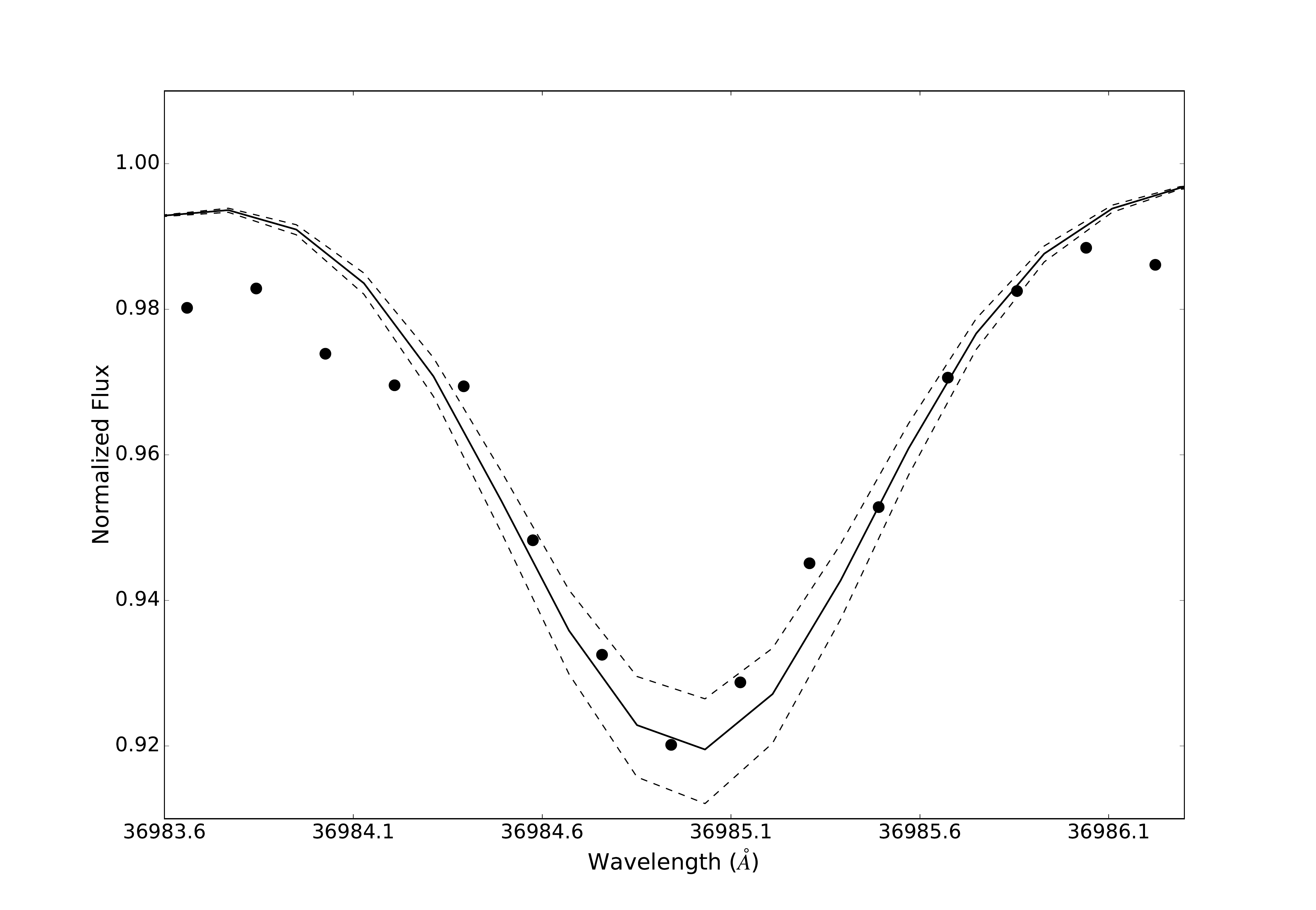} 
\caption{H$^{35}$Cl feature in RZ Ari. The solid line is the best fit determined by eye with our synthetic spectrum for A($^{35}$Cl) of 4.82 and the dashed lines show fits with $\pm$0.05 dex in A(Cl). The dots represent the observed RZ Ari spectrum. \label{fig::rzari_clfeature}} 
\end{figure*}

The spectral analysis program MOOG \citep{sneden} 2014 version, was used in conjunction with MARCS models \citep{gustafsson} to fit synthetic spectra to each star in Table \ref{table::obslog}. The synthetic spectra were convolved with a Gaussian function to match both the depth and wings of the spectral features. The fit to the entire spectrum of RZ Ari can be found in Figure \ref{fig::exspec2}. The upper panel of the figure shows an observed spectrum before telluric lines were removed and the middle panel shows only the telluric features contaminating the data. The fit to the H$^{35}$Cl feature for RZ Ari is shown in Figure \ref{fig::rzari_clfeature}. The abundances of C, N, O, Si, Cl, K, and Ca were adjusted to obtain the best fit determined by eye to the spectrum and the resulting abundances are included in Table \ref{table::params}. The spectrum of the M-dwarf BD+68 946 resembles the sunspot umbra shown in Figure \ref{fig::loggffits} and we were unable to derive abundances for  \ion{Ca}{1}, \ion{Si}{1}, and \ion{K}{1} without modeling the magnetic field and including the Zeeman components for each line.

Atmospheric models were created for seven stars that span the full metallictity and temperature range of the sample to estimate the abundance uncertainties due to model parameters. The uncertainties for each element were averaged and the total errors can be seen in Table \ref{table::uncertainties}. The average errors for each parameter were assumed independent of one another and added in quadrature. The average uncertainty was adopted for each star. 

Uncertainties on the fit were quantified for seven of the sample stars. By altering the abundances for each element to find fit the range of abundance that could reasonably fit each spectrum. The average uncertainties for each element can be found in Table \ref{table::uncertainties}. Carbon showed the most uncertainty associated with the fit; only one weak and blended CH line was present and reliable in each spectrum. Other CH lines were potentially identified but could not be fit to features in the solar and Arcturus spectra. Abundant molecules in cool atmospheres include NH, CH, OH, CN and CO. Therefore, a change to the carbon abundance will affect the amount of CO present in the star's atmosphere and will change the strength of observed OH lines. While the CH feature present in the spectra of these stars is weak, as seen in Figure \ref{fig::exspec2}, it can estimate the carbon abundance. 

The CH, NH, and OH lines also depend on all the abundances of the CNO trio. Lowering the oxygen abundance increases the strength of CH features due to the reducing the number of CO molecules and increasing the number of CH molecules. This resulted in 0.1-0.2 dex difference from the measured abundances and is added to the overall uncertainty. Nitrogen abundances could be derived from NH molecular lines found in our spectra. We found when the nitrogen abundance changed the strength of the CH and OH line features were not strongly affected;  the change in C or O abundance associated with altering the N abundance was less than the error from fitting the synthetic spectra the OH and CH to molecular features.

\begin{figure*}[t!]
\epsscale{1} 
\centering
\includegraphics[trim=0cm 0cm 0cm 0cm, scale=.35, clip=True]{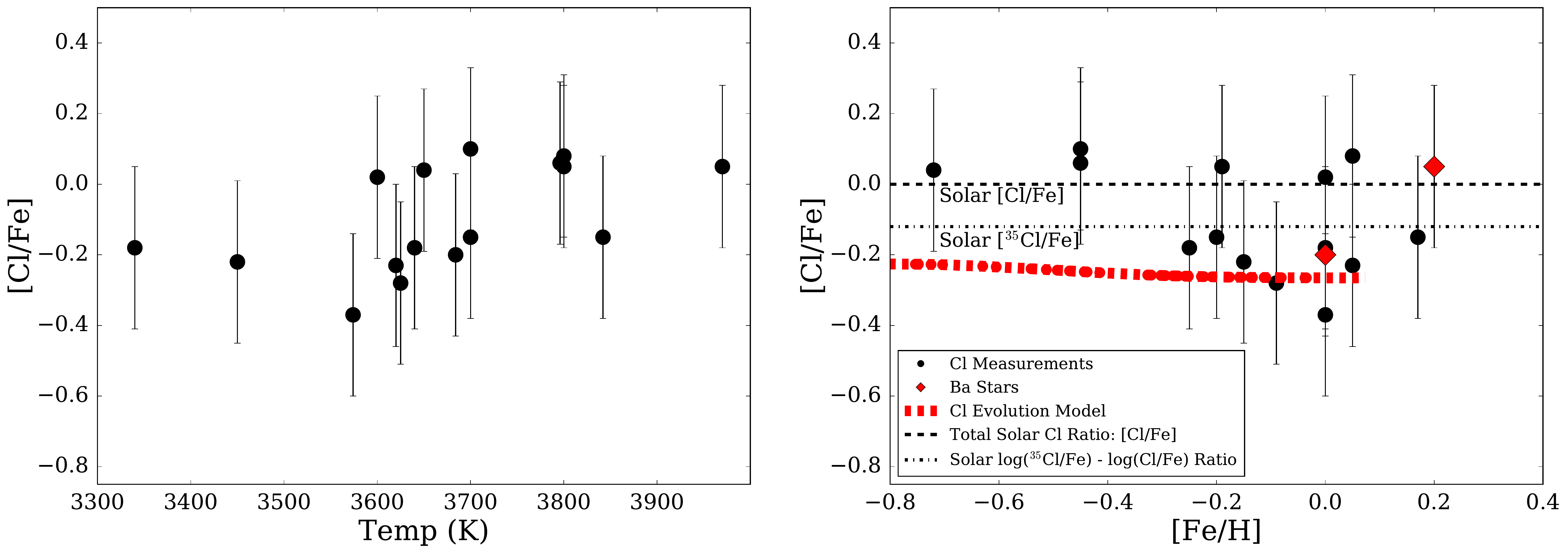} 
\caption{Teff and [$^{35}$Cl/Fe] versus temperature and [Fe/H] for our stars. The red dashed line represents a chemical evolution model for stars in the solar neighborhood \citep{kobayashi11}. Red diamonds represent barium stars and the open circle with a dot represents the solar Cl abundance. A dashed lines show the solar [Cl/Fe] value and the dot-dashed line shows what the log($^{35}$Cl/Fe) - log(Cl/Fe) for comparison to our sample stars.  \label{fig::results}} 
\end{figure*}

\begin{deluxetable*}{l c c c c c | c c c c c}
\tablewidth{0pt} 
\tablecaption{Abundance Comparisons \label{table::abundances_comparison}} 

\tablehead{ \multicolumn{6}{c}{This Paper} & \multicolumn{2}{c}{Literature Values} \\  \colhead{Star} & \colhead{A(C)} & \colhead{A(N)} & \colhead{A(O)} & \colhead{A(Si)} & \colhead{A(Ca)} & \colhead{A(C)} & \colhead{A(N)}& \colhead{A(O)} & \colhead{A(Si)} & \colhead{A(Ca)}} 
\startdata
HD 6860\tablenotemark{d} & 8.33  & 7.96  & 8.62 & 7.09 & 6.11 & 8.06 & 8.05 & 8.78 & 7.18 & 6.02 \\
HD 10380\tablenotemark{a} & 8.47 & 7.83 & 8.67 & 7.22 & 6.08 & \nodata & \nodata & \nodata & 7.34 & 6.03 \\
HD 18191\tablenotemark{e} & 8.33 & 7.99 & 8.47 & 7.36 &	 6.43 & 8.10& \nodata & 8.57 & \nodata & \nodata \\
HD 18884\tablenotemark{e} & 8.33 & 7.63 & 8.60 &  6.97 & 6.25  & 8.64 &  \nodata & 8.98 & & \nodata \\
HD 20468\tablenotemark{b} & 8.43 &  7.78 & 8.60 & 6.97 & 6.07 & 8.16 & \nodata & 8.56 & \nodata & \nodata \\
HD 30959\tablenotemark{c}  & 8.38 & 8.09 & 8.56 & 7.41 & 6.13 &  8.49 & 8.26 & 8.75 & \nodata & \nodata \\
HD 35155\tablenotemark{c} & 8.30 & 7.64 & 8.45 & 7.04 & 5.91  & 8.40 & 8.06 & 8.80 & \nodata & \nodata \\
HD 44478\tablenotemark{e} & 8.43 & 8.03 & 8.83 & 7.43 & 6.37 & 8.32 & \nodata & 8.81 & \nodata & \nodata \\
HD 49368\tablenotemark{c} & 8.42 & 7.99  & 8.56 & 7.26 & 6.24 & 8.43 &  8.50 & 8.80 & \nodata & \nodata \\
HD 112300\tablenotemark{e}  & 8.70 &  8.07 & 8.96 & 7.23 & 6.44 & 8.50 & \nodata & 8.84 & \nodata & \nodata  \\
HD 147923\tablenotemark{c}  & 8.44 & 8.19 & 8.64 & 7.25 & 6.25 & 8.39 & 8.70 & 8.74 & \nodata & \nodata\\
HD 216386\tablenotemark{e} & 8.61 & 8.17 & 8.92 & 7.26 & 6.21 & 8.56 & \nodata & 8.92 & \nodata & \nodata \\
\enddata
\tablenotetext{a}{\citet{hinkle14}}
\tablenotetext{b}{\citet{luck}}
\tablenotetext{c}{\citet{smith2}}
\tablenotetext{d}{\citet{smith}}
\tablenotetext{e}{\citet{tsuji}}
\end{deluxetable*}

\begin{figure*}[b!]
\epsscale{1} 
\centering
\includegraphics[trim=0cm 0cm 0cm 0cm, scale=.35, clip=True]{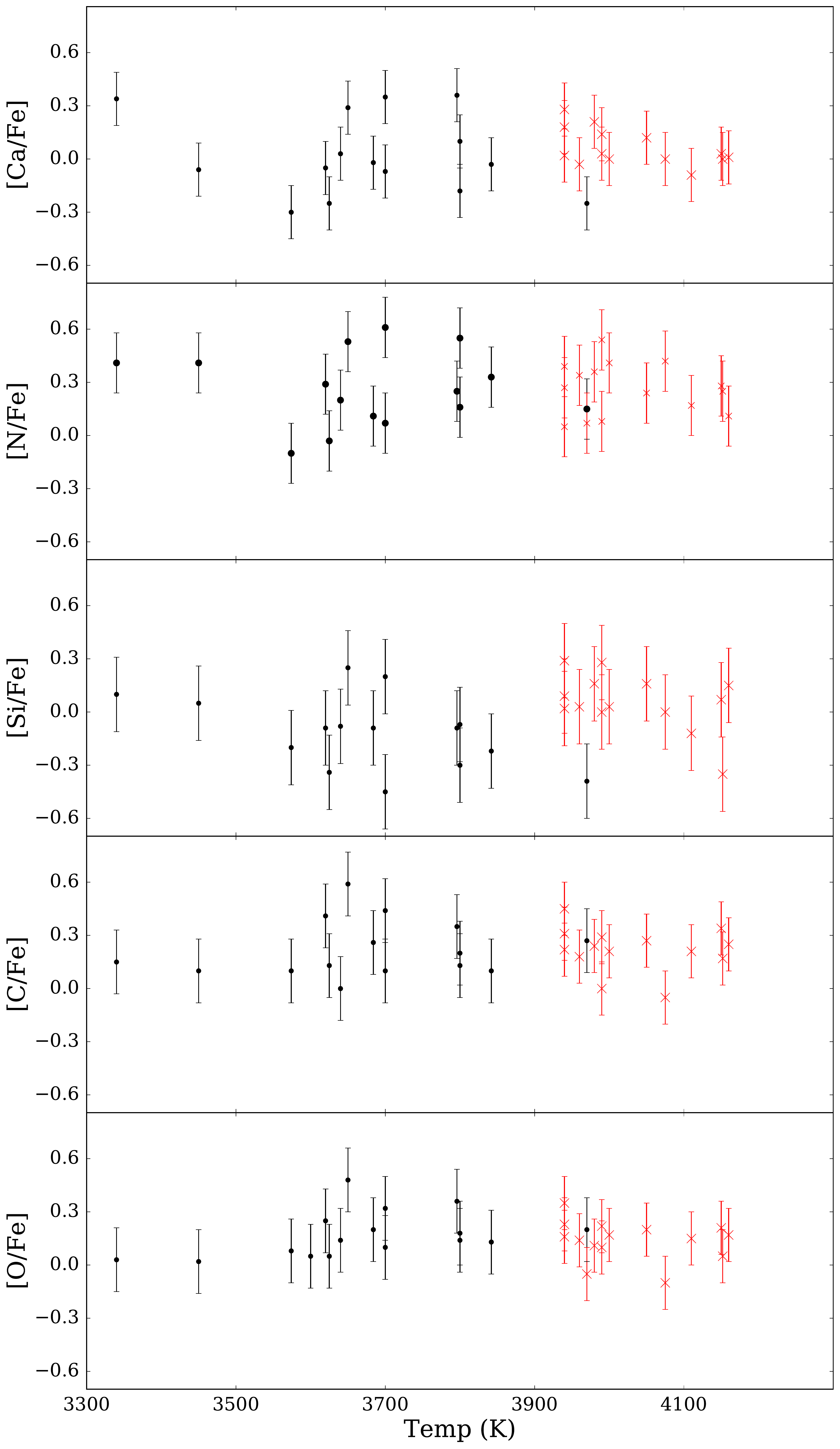} 
\caption{Temperature versus [X/Fe] abundance for each element in our sample. Black circles represent data that has measurable $^{35}$Cl features while the red crosses show stars with temperatures too high to produce HCl features.  \label{fig::temp}} 
\end{figure*}

The final [$^{35}$Cl/Fe] abundances and their temperature dependence are shown in Figure \ref{fig::results}. The slope of [$^{35}$Cl/Fe] versus T$_{eff}$ is flat, with no systematic dependence on temperature. Additionally, the chosen H$^{35}$Cl feature is only measurable in stars cooler than 3900 K with the exception of the metal rich star HD 138481. The dependence of the derived abundances for C, N, O, Si, and Ca on temperature is shown in Figure \ref{fig::temp}. The absence of any trend with temperature suggests our stellar effective temperatures are appropriate and that our atomic and molecular data are reliable. 

\subsection{C, N, O, Si, and Ca Abundance Comparisons}

\begin{figure*}[b!]
\epsscale{1} 
\centering
\includegraphics[trim=0cm 0cm 0cm 0cm, scale=.35, clip=True]{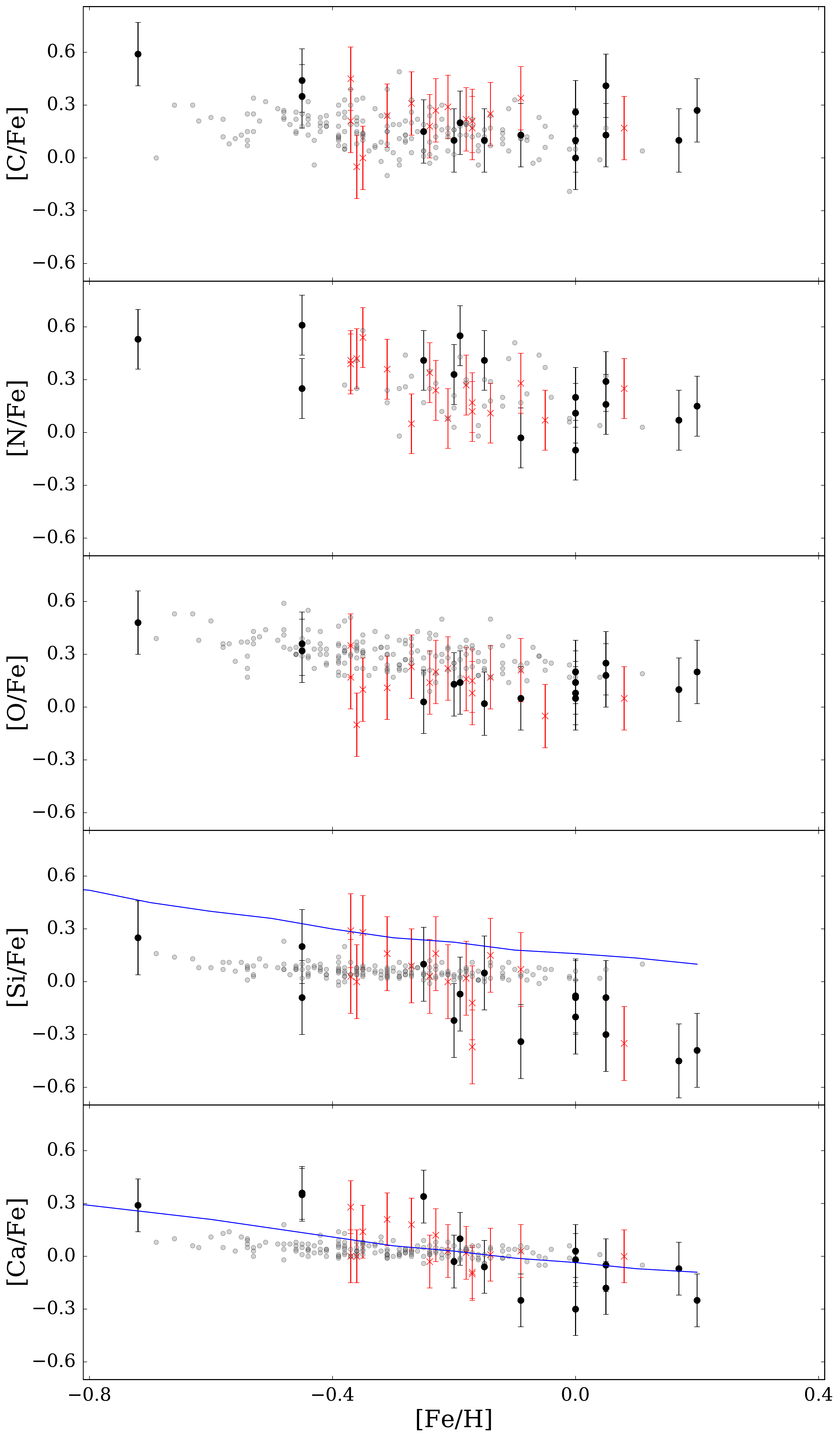} 
\caption{[X/Fe] versus [Fe/H] abundance for C, O, Si, and Ca. Filled circles represent data with measurable $^{35}$Cl abundances while the red crosses mark stars with temperatures too high to produce H$^{35}$Cl features. Grey filled points represent abundances from \citet{reddy}. The blue lines in the [Si/Fe] and [Ca/Fe] panels are chemical evolution models from fig. 10 in \citet{nomoto2}.  \label{fig::otherelements}} 
\end{figure*}

Our C, N, O, Si, and Ca abundances, plotted in Figure \ref{fig::otherelements}, are consistent with the literature for stars near solar metallicity. The average abundances we derived for the coolest stars in our sample are consistent with those for the warmest stars.

Additionally, abundances for C, N, O, Si, and Ca measured from features in our spectra can be compared with previous determinations for stars in our sample from the literature. The abundances derived from the L-band spectra match other near infrared and infrared measurements, shown in Table \ref{table::abundances_comparison}. Previous studies used equivalent width measurements of weak CO, CN, and OH features between 1-4 $\mu$m in infrared spectra \citep{smith, tsuji}. The absolute average difference between our carbon abundances and the literature values is 0.15 with a standard deviation of 0.09. Oxygen has an absolute average difference is 0.15 with a standard deviation of 0.12. Most stars therefore fall within our typical errors of 0.18 dex for both stars. Only five of our sample stars have previously measured nitrogen abundances. Two of our stars agree within one sigma while the three other stars  disagree by $\sim$0.5 dex.  The dispersion in carbon may be due to difficulties fitting our single CH feature. Three stars in our comparisons have oxygen abundances significantly larger than one standard deviation from the accepted literature values: HD 18884, HD 35155, and HD 49368. Additionally, both HD 35155, and HD 49368 are 2-3 sigma larger than the our measured abundances. These are the three lowest metallicity stars in our sample, and the discrepancy may be due to inaccurate initial metallicities. Follow up observations would be necessary to confirm the atmospheric parameters of these stars. Additionally, while the OH features are strong, we note that even the strongest equivalent width is log(EW/$\lambda$)$<$--4.77 

Our [x/Fe] ratio are compared against [Fe/H] for the program stars and F and G dwarfs in the thin disk \citep{reddy} in Figure \ref{fig::otherelements}. Our derived stellar abundances are similar to thin disk F and G dwarfs, further demonstrating the reliability of our abundances and temperatures derived in this spectral range. Again, the only significant discrepancy is with our oxygen abundances, which are generally $\sim$0.1 dex lower than typical F and G dwarf values. Additionally, Si abundances at high metallicity are lower than the F and G dwarfs in \citet{reddy}. Models of Ca and Si chemical evolution in the solar neighborhood (adopted from Fig. 10 in \citet{nomoto2}) are shown in Figure \ref{fig::otherelements}. Our abundances and the data from \citet{reddy} are consistent with the Ca model but fall below the \citet{nomoto2} model for Si. 

Oxygen abundances derived from infrared molecular OH lines are sensitive to 3-D hydrodynamical effects in stars where convection is an important energy flow \citep{kucinskas}. In a T$_{eff}$=3660 K, log(g)=1.00 [M/H]=0 star, \citet{kucinskas} found the abundance difference (3-D - 1-D) for OH lines with excitation potentials between 1-2 eV is --0.06 to --0.08 dex at 1.6 $\mu$m. This effect was noted to be less severe at increasing wavelength. Any 3-D hydrodynamical effects in our oxygen abundances should be $<$--0.10 dex. The weak molecular features, lack of any systematic change with temperature, and comparisons to the literature all confirm our photospheric C, N, O, Si, and Ca abundances.

\section{Discussion} \label{sec::discussion}

\subsection{HCl Formation and Sample}
Our analysis demonstrates that the chlorine abundance can be measured in solar metallicity stars with temperatures below 3900 K. Stars with temperatures above 3900 K are too warm to form the HCl molecule, with the one  exception of HD 138481, at an effective temperature of 3970 K but metal rich at [Fe/H]=+0.2 \citep{smith2}. The lowest metallicity star with detectable H$^{35}$Cl in our sample is HD 35155 at [Fe/H]=--0.72. 

The majority of stars with H$^{35}$Cl features in  our sample are evolved stars of spectral types M, K, and S. Late-type dwarfs are difficult to observe at 3.7 $\mu$m due to their faint L-band apparent magnitude but the M dwarf BD+68 964 was bright enough to be included. This star has a measurable H$^{35}$Cl with [O/Fe] and [$^{35}$Cl/Fe] abundances that are consistent with the abundances measured in the evolved giants. The K8 dwarf HD 88230 is also included in our sample but  H$^{35}$Cl is not measurable due this star's effective temperature of 3970 K. 

\begin{figure*}[t!]
\epsscale{1} 
\centering
\includegraphics[trim=3cm 0cm 3cm 0cm, scale=.38, clip=True]{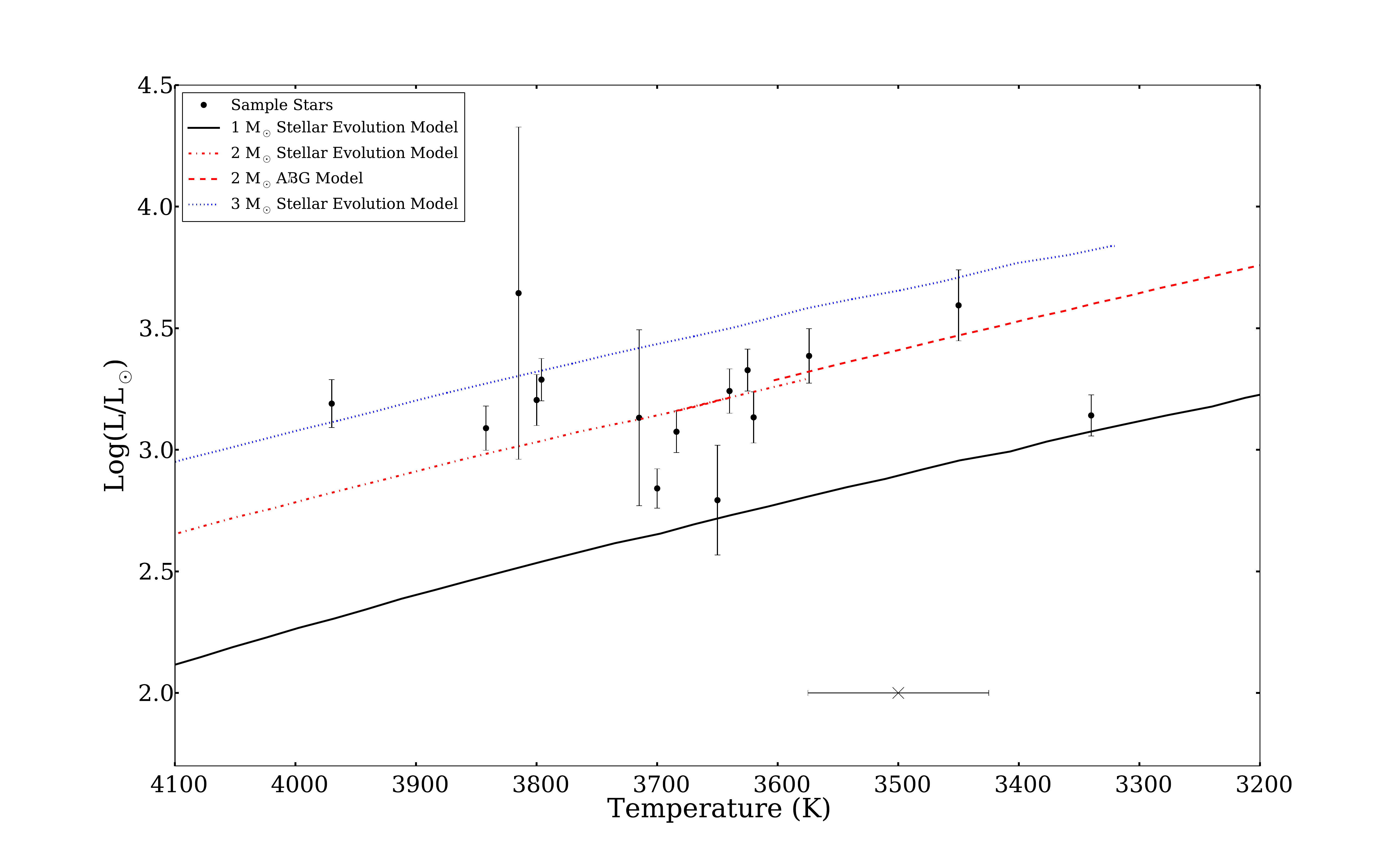} 
\caption{Stars with $^{35}$Cl abundances in a luminosity versus temperature plot. The lines show 1, 2, and 3 M$_{\odot}$ stellar evolutionary tracks from \citet{bertelli08, bertelli09}. The typical temperature uncertainty for the plotted stars is shown at the bottom of the figure. \label{fig::stellarevol}} 
\end{figure*}

The evolutionary state and masses of the stars in our sample with chlorine abundance measurements can be determined from absolute magnitudes computed using Hipparcos parallaxes \citep{leeuwen} and 2MASS infrared photometry \citep{skrutskie}. The 2MASS K$_{s}$ magnitude was converted to the Johnson K magnitude using methods of \citet{johnson}. A bolemetric correction for each star could then be determined from its J-K color and known atmospheric parameter \citep{bessell}. The input parameters and results for each star can be found in Table \ref{table::bolcor}.  Infrared photometry from IRAS \citep{beichman}, WISE \citep{wright}, and 2MASS  \citep{skrutskie} allowed the identification of any infrared excess due to circumstellar material around each star. Stars with possible infrared excesses were then compared to \citet{mcdonald}; only HD 147923 had J and K band magnitudes that were affected by more than 0.1 mags from the star's intrinsic magnitude \citep{mcdonald}. However, the distance to this star dominates the uncertainty on the luminosity rather than uncertainties in the J and K magnitudes. The stars' position in the Hertzsprung-Russel diagram are compared to models for a star with Z=0.017 and Y=0.30 \citep{bertelli08,bertelli09} and the results are plotted in Figure \ref{fig::stellarevol}. The stars in our sample have masses between 1 $M_{\odot}$ and 3 $M_{\odot}$ and are consistent with the locus of stars on the red giant branch. All but one star is located less than 500 pc from the Sun and the sample is therefore representative of stars in the solar neighborhood.

\begin{deluxetable*}{c c l l l l l l}
\tablewidth{0pt} 
\tablecaption{Distances and Luminosities \label{table::bolcor}} 
\tablehead{\colhead{Star} & \colhead{HD} & \colhead{Distance} & \colhead{K$_{s}$} & \colhead{(J-K$_{s}$)} & M$_{K}$ & BC$_{k}$\tablenotemark{a} & Log(L/$L_{\odot}$) \\
\colhead{} & \colhead{Number} & \colhead{(pc)} & \colhead{(Mag)} & \colhead{} & \colhead{mag} & \colhead{mag} & \colhead{}} 
\startdata
BD+68 946 & - & 4.53 $\pm$ 0.02 &  4.60 & 0.79 &  6.33 $\pm$ 0.01 & --2.51 & -1.64 $\pm$ 0.08\\
$\beta$ And& 6860 & 60.5 $\pm$ 3.0 & --1.85 & 0.89 & --5.69 $\pm$ 0.11 & --2.71 & 3.09 $\pm$ 0.09 \\
RZ Ari & 18191 & 107.8 $\pm$ 3.5 & --0.87 &	1.10 & --5.96 $\pm$ 0.07 & --2.85 & 3.14 $\pm$ 0.08 \\
$\alpha$ Cet & 18884 & 76.4 $\pm$ 3.0 & --1.82 &  1.10 & --6.17 $\pm$ 0.09 & --2.69 & 3.29 $\pm$ 0.09 \\
Omicron Ori & 30959 & 200$ \pm$ 28 & --0.66 & 1.15 & --7.10 $\pm$ 0.30 &  --2.85 & 3.60 $\pm$ 0.15 \\
V1261 Ori  & 35155 & 288 $\pm$ 70 & 2.14 & 1.20 & --5.10 $\pm$ 0.53 & --2.85 & 2.80 $\pm$ 0.23\\
$\mu$ Gem  & 44478 & 71.0 $\pm$ 3.6 & --1.86 & 0.95 & --6.05 $\pm$ 0.11 & --2.69 & 3.24 $\pm$ 0.09\\
V613 Mon & 49368 & 495 $\pm$ 201 & 2.46 &  1.30 & --5.94 $\pm$ 0.88 &  --2.85 & 3.13 $\pm$ 0.36 \\
27 Cnc & 71250 & 276 $\pm$ 25 & 0.58 & 0.99 & --6.57 $\pm$ 0.20 & --2.84 & 3.39 $\pm$ 0.11  \\
$\delta$ Vir & 112300 & 60.8 $\pm$ 0.8 & -1.19 & 1.08 & --5.04 $\pm$ 0.03 & --2.68 & 2.84 $\pm$ 0.08 \\
83 UMa & 119228 & 160.5 $\pm$ 5.7 & 0.34 & 1.03 & --5.63 $\pm$ 0.08 & --2.68 & 3.08 $\pm$ 0.09 \\
AW CVn & 120933 & 184 $\pm$ 6.8 & --0.01 & 1.05 & --6.26 $\pm$ 0.08 & --2.69 & 3.33 $\pm$ 0.09\\
52 Boo & 138481 & 257 $\pm$ 17 & 1.21 & 0.85 & --5.78 $\pm$ 0.14 & --2.54 & 3.19 $\pm$ 0.10\\
\nodata & 147923 & 1400 $\pm$ 1100 & 3.46 & 1.25 & --7.21 $\pm$ 1.70 & --2.84 & 3.64 $\pm$ 0.70\\
$\lambda$ Aqr & 216386 & 118.1 $\pm$ 9.2 & --0.67 & 1.09 & --5.96 $\pm$ 0.17 & --2.69 & 3.20 $\pm$ 0.10 \\
30 Psc (YY Psc) & 224935 & 132.5 $\pm$ 10.4 & --0.40 & 1.02 & --5.94 $\pm$ 0.17 & --2.85 & 3.13 $\pm$ 0.11 \\
\enddata
\tablenotetext{a}{Adopted uncertainty of $\pm$0.2 for all Bolemetric Corrections (BC$_{k}$)}
\end{deluxetable*}

\subsection{Chlorine Isotope Ratios}

The $^{35}$Cl/$^{37}$Cl isotope ratio for the solar system has been measured to be 3.13 \citep{lodders}. A survey of HCl in the Galaxy from molecular clouds, protoplanetary cores, and evolved stars (such as IRC+10216) revealed isotope ratios between 1$<^{35}$Cl/$^{37}$Cl$<$5 with most values between 1 and 3 \citep{peng}. The range of values is attributed to different isotope production in Type Ia and CCSNe \citep{peng}. Type Ia supernova produce yield ratios of $^{35}$Cl/$^{37}$Cl $\sim$ 3.5 to 5.5 depending on the model parameters, such as ignition, 2D versus 3D models, and model resolution \citep{travaglio}. CCSNe models produces chlorine yields with isotope ratios of 2.32 for Z=0.004, 18 M$_{\odot}$ stars, 1.2 for Z=0.02, 25 M$_{\odot}$ stars, and  1.75 for high energy supernova with Z=0.02 and 25 M$_{\odot}$ \citep{kobayashi11}.   

Our sample includes only one star, RZ Ari with a measurable H$^{37}$Cl feature at 3.7010 $\mu$m, as shown in Fig. \ref{fig::exspec2}. Other stars, including the sunspot umbral spectra, were too warm to form an H$^{37}$Cl feature significantly above the noise level; the H$^{37}$Cl feature is stronger at both lower temperature and higher abundance. The isotope ratio in RZ Ari was determined by comparing the equivalent widths of the two HCl features. The equivalent widths are EW(H$^{35}$Cl)=(81 $\pm$ 6) m$\AA$ and EW(H$^{37}$Cl)=(36 $\pm$ 6) m$\AA$. Uncertainties are found from the standard deviation of multiple measurements of the equivalent width at different continuum values. The final isotopic abundance ratio is $^{35}$Cl/$^{37}$Cl = 2.2 $\pm$ 0.4. This value was also confirmed by spectral synthesis; this isotope ratio was used to fit the H$^{37}$Cl feature, shown in Figure \ref{fig::exspec2}, with the best fit determined by eye. 

The isotope ratio for Cl for different metallicities in the solar neighborhood has been predicted by \citet{kobayashi11}: Cl$^{35}$/Cl$^{37}$ = 1.79 at [Fe/H]=0, and 1.94 at [Fe/H] = --0.5. RZ Ari has an [Fe/H] of --0.25. Our measured chlorine isotope value of 2.2 $\pm$ 0.4 is near one standard deviation of these predicted values. This ratio is also consistent with the range of isotope measurements found in the interstellar medium \citep{peng} and is less than two standard deviations from the solar isotopic ratio. Further measurements in cool stars are necessary to explore the dispersion of the isotope ratio in the local solar neighborhood. 

\subsection{Chlorine in Planetary Nebulae and \ion{H}{2} Regions}

\begin{figure*}[t!]
\epsscale{1} 
\centering
\includegraphics[trim=0cm 0cm 0cm 0cm, scale=.32, clip=True]{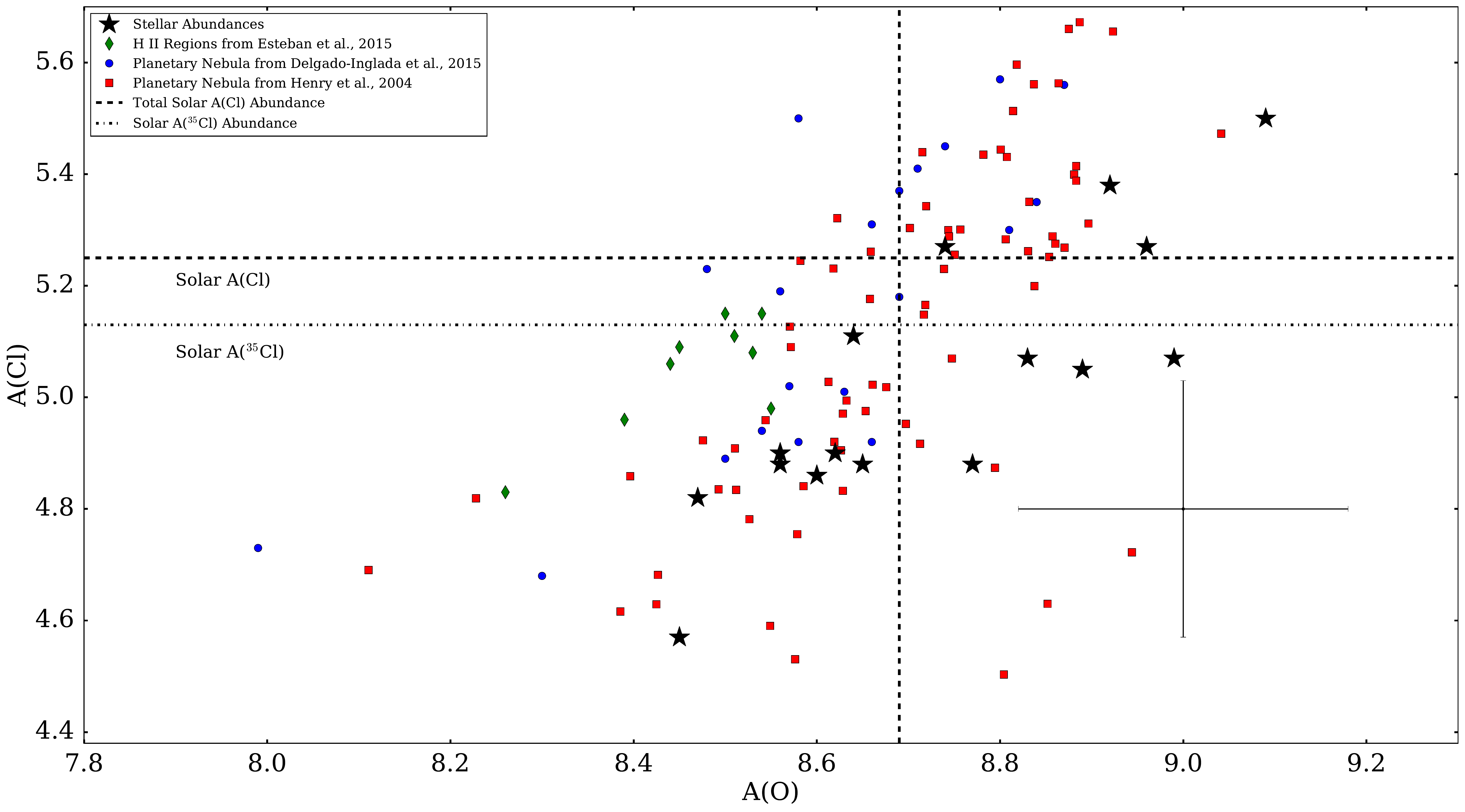} 
\caption{Abundance of A(Cl) versus A(O) in multiple Galactic sources. The black stars represent A($^{35}$Cl) stellar abundances from this sample. Diamonds are Cl abundances in \ion{H}{2} regions from \citet{esteban}. Red squares and blue circles are Cl abundances from planetary nebula from \citet{henry} and \citet{delgado} respectively. The dashed lines indicate the solar oxygen abundance of A(O)=8.69 and total chlorine abundance of A(Cl)=5.25. The dot-dashed line represents the solar A($^{35}$Cl)=5.13 abundance.\label{fig::cl_vs_o}} 
\end{figure*}

Figure \ref{fig::cl_vs_o} compares the $^{35}$Cl abundance in our sample of stars with the total Cl abundance measured in Galactic planetary nebula and \ion{H}{2} regions. Studies of planetary nebula have found the radial Galactic abundance gradients of Cl and O to be statistically indistinguishable from one another \citep{henry} and studies of Cl in \ion{H}{2} regions found O and Cl had identical radial Galactic abundance gradients of --0.043 dex kpc$^{-1}$ \citep{esteban}. This indicates both elements, Cl and O, are produced in lockstep evolution \citep{henry}. 

Our sample of stars cover too narrow a range of galacto-centric distance to allow us to determine a gradient, but the average A($^{35}$Cl) of our sample is consistent with what is predicted for the solar neighborhood by \citet{henry} and \citet{esteban}.  The offset between the planetary nebula and stellar abundances shown in Figure \ref{fig::cl_vs_o} is estimated to be 0.16 dex. This offset is likely due to the difference between the total Cl abundance reported in the atomic planetary nebula lines and $^{35}$Cl abundances measured from the H$^{35}$Cl molecular feature in our sample. For example the A($^{35}$Cl) abundance in RZ Ari is 4.82 and the full abundance is A(Cl)=4.98 for an isotope ratio of $^{35}$Cl/$^{37}$Cl of 2.2, an increase of 0.16 dex. Additionally, the solar isotope ratio is 3.13 which gives a $^{35}$Cl abundance in the sun of 5.13, a difference of 0.12 dex between the total abundance and $^{35}$Cl abundance. Therefore while our results show scatter, the stellar abundances follow similar Cl to O ratios to those seen in the nebular sources. Future observations will use the methodology in this paper to determine Cl isotope ratios in a larger sample of cool stars. 

\subsection{Chlorine and Chemical Evolution}

We present the first direct measurements of chlorine in stellar sources, where Cl can be directly compared to the iron abundance. Figure \ref{fig::results} shows the [$^{35}$Cl/Fe]\footnote{[$^{35}$Cl/Fe]=A($^{35}$Cl)-A(Cl)$_{\odot}$+[Fe/H]} ratio measured in our stellar sample versus a chemical evolution model from \citet{kobayashi11}. The average chlorine abundance of our sample is [$^{35}$Cl/Fe]=--0.10 $\pm$ 0.14. The chemical evolution model predicts that [$^{35}$Cl/Fe] to be nearly constant with metallicity, declining by only 0.03 dex from [Fe/H]=--0.6 to solar metallicity. Our [$^{35}$Cl/Fe] ratios are on average higher than the model by 0.16 $\pm$ 0.15 dex. The lowest metallicity stars have an average [$^{35}$Cl/Fe] ratio of 0.1 dex, about 0.2 dex higher than the solar metallicity but the sample contains too few metal-poor stars to conclude whether or not [$^{35}$Cl/Fe] declines with decreasing metallicity. Despite the small offset our results are still nearly consistent abundance of chlorine predicted for the solar neighborhood by chemical evolution models. 

\begin{figure*}[t!]
\epsscale{1} 
\centering
\includegraphics[trim=0cm 0cm 0cm 0cm, scale=.33, clip=True]{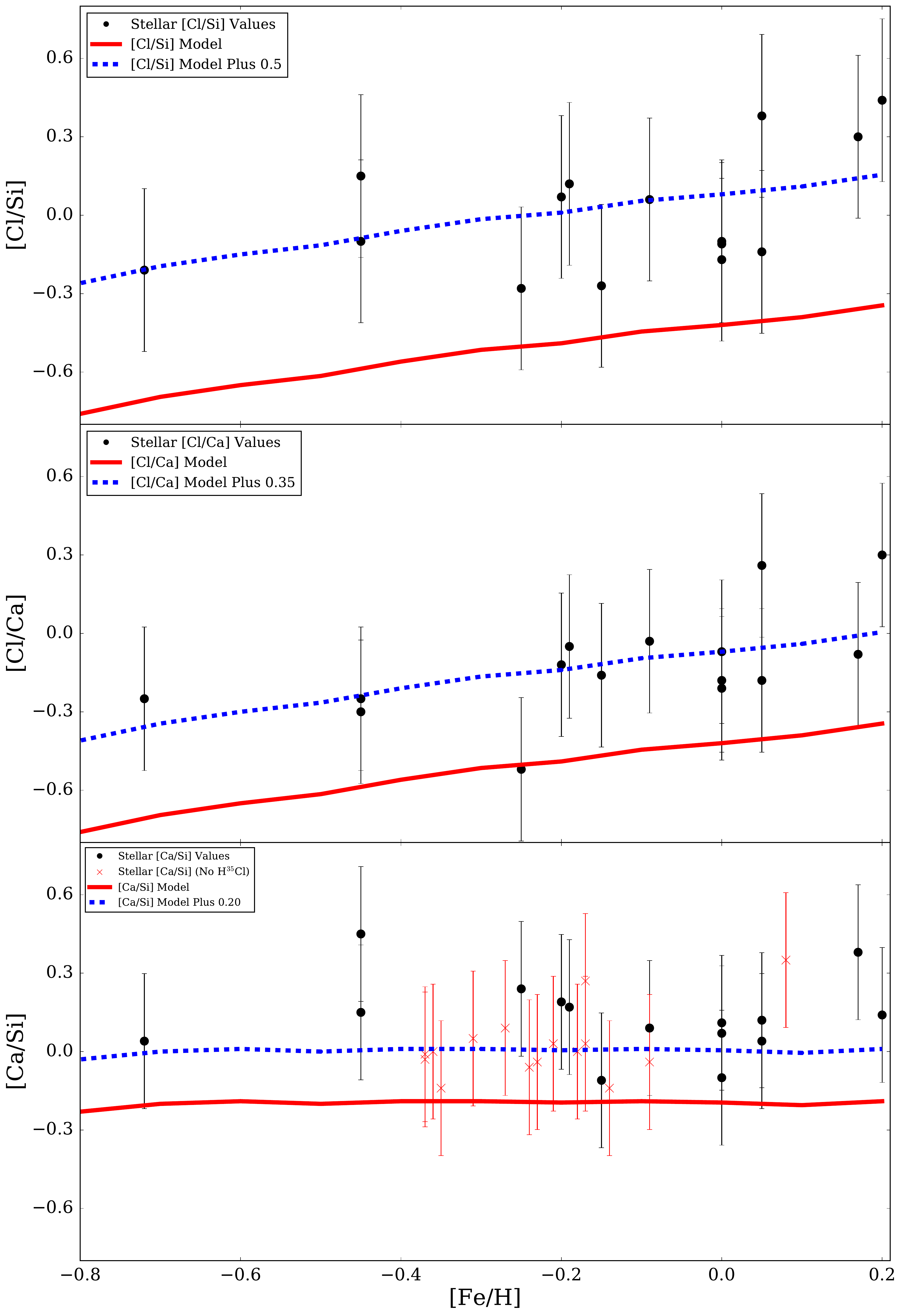} 
\caption{[Ca/Si], [$^{35}$Cl/Ca] and [$^{35}$Cl/Si] for our sample are plotted versus [Fe/H]. Solid red lines are chemical evolution models for the solar neighborhood, reproduced from figure 10 in \citet{nomoto2}. The blue dashed lines are the chemical evolution models with a constant offset added to their values.    \label{fig::cl_alpha}} 
\end{figure*}

Chlorine abundances were also compared with the alpha elements silicon and calcium derived from our spectra. Figure \ref{fig::cl_alpha} shows our observed abundance ratios versus chemical evolution models (reproduced from Figure 10) from \citet{nomoto2}. Our measured [$^{35}$Cl/Si] and [$^{35}$Cl/Ca] ratios are higher than predicted by \citep{nomoto} by typically 0.4 dex. Offsetting each chemical evolution model to match the observed abundance, as shown in Fig. 10, demonstrate that the slope of the fit is consistent with the measured abundances and that the discrepancies do not depend on metallicity in our sample. The bottom panel in Figure \ref{fig::cl_alpha} shows the [Ca/Si] ratio is also offset compared to the model value. The [Ca/Si] offset was found to be 0.20 dex. 

We find the [$^{35}$Cl/Ca] ratio increases at higher [Fe/H]. This is expected because [$^{35}$Cl/Fe] is predicted to be constant with metallicity while [Ca/Fe] decreases as metallicity increases. The offset between models and chlorine abundance may be due to the additional production of Cl by the $\nu$ process, which may affect yields of Cl \citep{kobayashi15} and is not included in the present model. These offsets between the chemical evolution models and the observed abundances put constraints on the strength of any additional process that produces Cl. 

We are also comparing only $^{35}$Cl abundances to the Cl chemical evolution models. Including the $^{37}$Cl abundance would increase the overall abundance by $\sim$0.1 to 0.2 dex depending on the isotope ratio. For example the A($^{35}$Cl) abundance in RZ Ari is 4.82 and the full abundance is A(Cl)=4.98 for an isotope ratio of $^{35}$Cl/$^{37}$Cl of 2.2.

To determine if chlorine is produced using the s-process, two low temperature barium-enriched stars were included in our sample. HD 138481 is a class 0.5 Ba star and HD 119228 is a class 0.5 Ba star \citep{lu}, represented as red diamonds in Figure \ref{fig::results}. The [$^{35}$Cl/Fe] in those stars is similar to normal giants and suggests that additional Cl production via the s-process is unlikely. 

\section{Conclusions}

We used L-band infrared spectroscopy to measure the abundance of chlorine in stars with effective temperatures below 3900 K, with the exception of the metal-rich HD 138481 which has a temperature of 3970 K. Theoretical log gf values were used and compared to astrophysical log gf values derived from fitting synthetic spectra to Arcturus, a quiet-sun photosphere spectrum, and a solar sunspot umbrae spectrum. Of the sample, 16 stars showed measurable Cl features in their spectra. The main results from the abundance analysis are summarized below.

\begin{enumerate}

\item{Our sample consisted of M and K giants and dwarfs. The H$^{35}$Cl molecular feature is detected in 15 giants and one M-dwarf, all with effective temperatures below 3900 K except for HD 138481 which has a temperature of 3970 K. The masses of the evolved stars are determined from stellar evolution models and are found to be between 1 and 3 M$_{\odot}$  }

\item{Chlorine and oxygen abundances in stars are consistent with measurements made in planetary nebulae \citep{henry, delgado} and \ion{H}{2} regions \citep{esteban}.}

\item{A $^{35}$Cl/$^{37}$Cl isotope ratio of 2.2$\pm$0.4 is found in RZ Ari. This result is consistent with measurements in the interstellar medium \citep{peng} and is near the solar value of 3.13 \citep{lodders}. It is also comparable to predicted isotope ratios of $\sim$ 1.8-1.9 from supernova yields \citep{kobayashi11}}

\item{Abundance measurements are determined for C, N, O, Si, and Ca are consistent with similar abundances in warmer stars in our sample ($\sim$ 4300 K). The abundances from our L-band spectra also matched abundances determined previously for these stars from the literature. Our abundances also matched F and G dwarfs in the solar neighborhood \citep{reddy}. }

\item{[$^{35}$Cl/Fe] measurements are consistent with the slope of Galactic chemical evolution models from \citet{kobayashi11} but show on average a higher [$^{35}$Cl/Fe] abundances than predicted by 0.16 dex. This offset is also seen in [$^{35}$Cl/Ca] and [$^{35}$Cl/Si] where chlorine is offset by $\sim$0.35 dex compared to chemical evolution models \citep{nomoto2}. An additional processes producing chlorine may be necessary to explain the overabundance of Cl compared to predictions. The $\nu$ process may affect Cl production \citep{kobayashi15} and should be considered. }

\item{Chlorine abundances in two Ba-rich stars are similar to the chlorine abundance in other, non s-process rich stars in the sample. The similarity of these chlorine abundances suggests that additional A($^{35}$Cl) is not produced through the s-process}

\end{enumerate}

\section{Acknowledgments}
We thank the anonymous referee for their thoughtful comments and suggestions on the manuscript. This paper is based on observations obtained at the Kitt Peak National Observatory, a division of the National Optical Astronomy Observatories (NOAO). NOAO is operated by the Association of Universities for Research in Astronomy, Inc. under cooperative agreement with the National Science Foundation. We are grateful to the Kitt Peak National Observatory and particularly to Colette Salyk for her assistance at the start of the observing run and to Anthony Paat, Krissy Reetz, and Doug Williams during the run. This research has made use of the NASA Astrophysics Data System Bibliographic Services, the HITRAN database and the Kurucz atomic line database operated by the Center for Astrophysics, and the Vienna Atomic Line Database operated at the Institute for Astronomy of the University of Vienna. This research has made use of the SIMBAD database, operated at CDS, Strasbourg, France. We thank Eric Ost for implementing the model atmosphere interpolation code.  C. A. P. acknowledges the generosity of the Kirkwood Research Fund at Indiana University.


\newpage

\begin{thebibliography}{}

\bibitem[Asplund et al.(2009)]{asplund} 
Asplund, M., Grevesse, N., Sauval, A.~J., \& Scott, P.\ 2009, \araa, 47, 481 
	
\bibitem[Beichman et al.(1988)]{beichman} Beichman, C.~A., Neugebauer, G., Habing, H.~J., Clegg, P.~E., \& Chester, T.~J.\ 1988, Infrared astronomical satellite (IRAS) catalogs and atlases.~Volume 1: Explanatory supplement, 1,  

\bibitem[Bernath (2015)]{bernath}
Bernath, P. F., 2016, Spectra of Atoms and Molecules, (3rd ed.; Oxford: Oxford University Press)

\bibitem[Bertelli et al.(2008)]{bertelli08} 
Bertelli, G., Girardi, L., Marigo, P., \& Nasi, E.\ 2008, \aap, 484, 815 

\bibitem[Bertelli et al.(2009)]{bertelli09}  
Bertelli, G., Nasi, E., Girardi, L., \& Marigo, P.\ 2009, \aap, 508, 355

\bibitem[Bessell et al.(1998)]{bessell} 
Bessell, M.~S., Castelli, F., \& Plez, B.\ 1998, \aap, 333, 231 

\bibitem[Brooke et al.(2014)]{brookenh}
Brooke J. S. A., Bernath P. F., Western C. M., van Hermert M. C., \& Groenenboom G. C., 2014, J. Chem. Phys., 141, 054310

\bibitem[Brooke et al.(2015)]{brookenh2}
Brooke J. S. A., Bernath P. F., \& Western C. M., 2015, J. Chem. Phys., 143, 026101

\bibitem[Brooke et al.(2016)]{brooke}
Brooke J. S.A., Bernath P. F., Western C. M., et al., 2016, J. Quant. Spec. Rad. Trans., 168, 142

\bibitem[Cenarro et al.(2007)]{cenarro} 
Cenarro, A.~J., Peletier, R.~F., S{\'a}nchez-Bl{\'a}zquez, P., et al.\ 2007, \mnras, 374, 
664 

\bibitem[Chen et al.(1998)]{chen} 
Chen, P.~S., Wang, X.~H., \& Xiong, G.~Z.\ 1998, \aap, 333, 613 

\bibitem[Delgado-Inglada et al.(2015)]{delgado}
Delgado-Inglada, G., Rodr{\'{\i}}guez, M., Peimbert, M., Stasi{\'n}ska, G., 
\& Morisset, C.\ 2015, \mnras, 449, 1797 

\bibitem[Ducati(2002)]{ducati}
Ducati, J.~R.\ 2002, VizieR Online Data Catalog, 2237,  

\bibitem[Espinosa-Garc{\'{\i}}a et al.(1995)]{espinosa} 
 Espinosa-Garc{\'{\i}}a, Corchado J. C., Fern{\'a}ndez J., \& Marquez A., 1995, Chem. Phys. Lett. 233, 220 

\bibitem[Esteban et al.(2015)]{esteban} 
Esteban, C., Garc{\'{\i}}a-Rojas, J., \& P{\'e}rez-Mesa, V.\ 2015, \mnras, 452, 1553 

\bibitem[Garc{\'{\i}}a-Rojas \& Esteban(2007)]{garcia} 
Garc{\'{\i}}a-Rojas, J., \& Esteban, C.\ 2007, \apj, 670, 457 

\bibitem[Gonz{\'a}lez Hern{\'a}ndez \& Bonifacio(2009)]{gonzalez} 
Gonz{\'a}lez Hern{\'a}ndez, J.~I., \& Bonifacio, P.\ 2009, \aap, 497, 497 

\bibitem[Gromadzki et al.(2013)]{gromadzki}
Gromadzki, M., Miko{\l}ajewska, J., \& Soszy{\'n}ski, I.\ 2013, \actaa, 63, 405 

\bibitem[Gustafsson et al.(2008)]{gustafsson} 
Gustafsson, B., Edvardsson, B., Eriksson, K., et al.\ 2008, \aap, 486, 951

\bibitem[Hall \& Noyes(1972)]{hall} 
Hall, D.~N.~B., \& Noyes, R.~W.\ 1972, \apjl, 175, L95 

\bibitem[Hekker \& Mel{\'e}ndez(2007)]{hekker} 
Hekker, S., \& Mel{\'e}ndez, J.\ 2007, \aap, 475, 1003 

\bibitem[Henry et al.(2004)]{henry} 
Henry, R.~B.~C., Kwitter, K.~B., \& Balick, B.\ 2004, \aj, 127, 2284 

\bibitem[Hinkel et al.(2014)]{hinkle14}
Hinkel, N.~R., Timmes, F.~X., Young, P.~A., Pagano, M.~D., \& Turnbull, M.~C.\ 2014, \aj, 148, 54 

\bibitem[Hinkle et al.(1995)]{hinkle} 
Hinkle, K., Wallace, L., \& Livingston, W.\ 1995, \pasp, 107, 1042 

\bibitem[Hinkle et al.(1998)]{hinkle_et_al_1998}
Hinkle, K. H., Cuberly, R., Gaughan, N., et al. 1998, Proc. SPIE, 3354, 810

\bibitem[Holtzman et al.(2015)]{holtzman} 
Holtzman, J.~A., Shetrone, M., Johnson, J.~A., et al.\ 2015, \aj, 150, 148 

\bibitem[Jofr{\'e} et al.(2014)]{jofre}
Jofr{\'e}, P., Heiter, U., Soubiran, C., et al.\ 2014, \aap, 564, A133

\bibitem[Johnson et al.(2005)]{johnson} 
Johnson, C.~I., Kraft, R.~P., Pilachowski, C.~A., et al.\ 2005, \pasp, 117, 1308 

\bibitem[Joyce(1992)]{joyce_1992}
Joyce, R. R. 1992, in ASP Conf. Ser. 23, Astronomical CCD Observing and
Reduction Techniques, ed. S. Howell (San Francisco: ASP), 258

\bibitem[Kama et al.(2015)]{kama}
Kama, M., Caux, E., L{\'o}pez-Sepulcre, A., et al.\ 2015, \aap, 574, A107

\bibitem[Kobayashi et al.(2006)]{kobayashi6} 
Kobayashi, C., Umeda, H., Nomoto, K., Tominaga, N., \& Ohkubo, T.\ 2006, \apj, 653, 1145

\bibitem[Kobayashi et al.(2011)]{kobayashi11}
Kobayashi, C., Karakas, A.~I., \& Umeda, H.\ 2011, \mnras, 414, 3231 

\bibitem[Kobayashi (2015)]{kobayashi15}
Kobayashi, C., 2015, Private Communication

\bibitem[Ku{\v c}inskas et al.(2013)]{kucinskas} 
Ku{\v c}inskas, A., Steffen, M., Ludwig, H.-G., et al.\ 2013, \aap, 549, A14 

\bibitem[Kumar et al.(1998)]{kumar}
A. Kumar, C. Hsiao1, W. Hung1  \& Y. Lee, 1998, J. Phys. Chem, 109, No. 10, 3824

\bibitem[Lambert et al.(1971)] {lambert} 
Lambert, D.~L., Mallia, E.~A., \& Brault, J.\ 1971, \solphys, 19, 289 

\bibitem[Lebzelter et al.(2001)]{lebzelter} 
Lebzelter, T., Hinkle, K.~H., \& Aringer, B.\ 2001, \aap, 377, 617 

\bibitem[Livingston \& Wallace(1991)]{livingston}
Livingston, W., \& Wallace, L.\ 1991, NSO Technical Report, Tucson: National Solar Observatory, National Optical Astronomy Observatory, 1991,  

\bibitem[Lodders et al.(2009)]{lodders} 
Lodders, K., Palme, H., \& Gail, H.-P.\ 2009, Landolt B{\"o}rnstein,  

\bibitem[Lu(1991)]{lu} 
Lu, P.~K.\ 1991, \aj, 101, 2229 

\bibitem[Luck \& Challener(1995)]{luck95}
Luck, R.~E., \& Challener, S.~L.\ 1995, \aj, 110, 2968

\bibitem[Luck(2014)]{luck} 
Luck, R.~E.\ 2014, \aj, 147, 137

\bibitem[Maiorca et al.(2014)]{maiorca}
Maiorca, E., Uitenbroek, H., Uttenthaler, S., et al.\ 2014, \apj, 788, 149 

\bibitem[Martin $\&$ Hepburn (1998)]{martin}
Martin J. D. D.  $\&$  Hepburn J. W., 1998, J. Phys. Chem., 109, No. 15, 8139

\bibitem[Masseron (2006)]{masseron} 
Masseron, T. 2006, PhD thesis, Obs. de Paris

\bibitem[Masseron et al.(2014)]{masseron_ch} 
Masseron, T., Plez, B., Van Eck, S., et al.\ 2014, \aap, 571, A47 

\bibitem[McDonald et al.(2012)]{mcdonald} 
McDonald, I., Zijlstra, A.~A., \& Boyer, M.~L.\ 2012, \mnras, 427, 343

\bibitem[McWilliam(1990)]{mcwilliam} 
McWilliam, A.\ 1990, \apjs, 74, 1075 

\bibitem[Mel{\'e}ndez et al.(2008)]{melendez}
Mel{\'e}ndez, J., Asplund, M., Alves-Brito, A., et al.\ 2008, \aap, 484, L21

\bibitem[Muller et al.(2014)]{muller}
Muller, S., Black, J.~H., Gu{\'e}lin, M., et al.\ 2014, \aap, 566, L6 

\bibitem[Nault(2014)]{nault} 
Nault K., 2014, Masters Thesis, IU

\bibitem[Nomoto et al.(2013)]{nomoto2}
Nomoto, K., Kobayashi, C., \& Tominaga, N.\ 2013, \araa, 51, 457

\bibitem[Nomoto et al.(2006)]{nomoto} 
Nomoto, K., Tominaga, N., Umeda, H., Kobayashi, C., \& Maeda, K.\ 2006, Nuclear Physics A, 777, 424

\bibitem[Peng et al.(2010)]{peng} 
Peng, R., Yoshida, H., Chamberlin, R.~A., et al.\ 2010, \apj, 723, 218 

\bibitem[Percy et al.(2008)]{percy} 
Percy, J.~R., Mashintsova, M., Nasui, C.~O., et al.\ 2008, \pasp, 120, 523 

\bibitem[Phillips(2007)]{phillips} 
Phillips, J.~P.\ 2007, \mnras, 376, 1120 

\bibitem[Prugniel et al.(2011)]{prugniel} 
Prugniel, P., Vauglin, I., \& Koleva, M.\ 2011, \aap, 531, A165 

\bibitem[Ram{\'{\i}}rez \& Allende Prieto(2011)]{ramirez}
Ram{\'{\i}}rez, I., \& Allende Prieto, C.\ 2011, \apj, 743, 135 

\bibitem[Reddy et al.(2003)]{reddy} 
Reddy, B.~E., Tomkin, J., Lambert, D.~L., \& Allende Prieto, C.\ 2003, \mnras, 340, 304

\bibitem[Rothman et al.(2013)]{rothman} 
Rothman, L. S., Gordon, I. E., Babikov, Y., et al. 2013, JQSRT, 130, 4

\bibitem[Ruscic et al.(2002)]{ruscic}
Ruscic B., Wagner F., Harding L., et al., 2002, J. Phys. Chem., 106, No. 11, 2727-2742

\bibitem[Sheffield et al.(2012)]{sheffield} 
Sheffield, A.~A., Majewski, S.~R., Johnston, K.~V., et al.\ 2012, \apj, 761, 161

\bibitem[Skrutskie et al.(2006)]{skrutskie} 
Skrutskie, M.~F., Cutri, R.~M., Stiening, R., et al.\ 2006, \aj, 131, 1163 

\bibitem[Smith \& Lambert(1990)]{smith2}
Smith, V.~V., \& Lambert, D.~L.\ 1990, \apjs, 72, 387 

\bibitem[Smith et al.(2013)]{smith} 
Smith, V.~V., Cunha, K., Shetrone, M.~D., et al.\ 2013, \apj, 765, 16 

\bibitem[Sneden(1973)]{sneden} 
Sneden, C.\ 1973, \apj, 184, 839

\bibitem[Soubiran et al.(2008)]{soubiran}
Soubiran, C., Bienaym{\'e}, O., Mishenina, T.~V., \& Kovtyukh, V.~V.\ 2008, \aap, 480, 91 

\bibitem[Sylwester et al.(2011)]{sylwester} 
Sylwester, B., Phillips, K.~J.~H., Sylwester, J., \& Kuznetsov, V.~D.\ 2011, \apj, 738, 49 

\bibitem[Tabur et al.(2009)]{tabur} 
Tabur, V., Bedding, T.~R., Kiss, L.~L., et al.\ 2009, \mnras, 400, 1945 

\bibitem[Travaglio et al.(2004)]{travaglio} 
Travaglio, C., Hillebrandt, W., Reinecke, M., \& Thielemann, F.-K.\ 2004, \aap, 425, 1029

\bibitem[Tsuji(2008)]{tsuji}
Tsuji, T.\ 2008, \aap, 489, 1271 

\bibitem[van Leeuwen(2007)]{leeuwen}
van Leeuwen, F.\ 2007, \aap, 474, 653 

\bibitem[Wallace et al.(2002)]{wallace2}
Wallace, L., Hinkle, K., \& Livingston, W.~C.\ 2002, NSO Technical Report \#02-001; Tucson: National Solar Observatory, 2002, 02 

\bibitem[Wright et al.(2010)]{wright} 
Wright, E.~L., Eisenhardt, P.~R.~M., Mainzer, A.~K., et al.\ 2010, \aj, 140, 1868-1881 

\bibitem[Worthey \& Lee(2011)]{worthey}
Worthey, G., \& Lee, H.-c.\ 2011, \apjs, 193, 1 

\bibitem[Woolf \& Wallerstein(2005)]{woolf} 
Woolf, V.~M., \& Wallerstein, G.\ 2005, \mnras, 356, 963

\bibitem[Woosley \& Weaver(1995)]{woosley} 
Woosley, S.~E., \& Weaver, T.~A.\ 1995, \apjs, 101, 181 

\end{thebibliography}
\end{document}